\documentclass[12pt]{article}

\usepackage{amsfonts}
\usepackage{amsmath}
\usepackage{amsthm}
\usepackage{cite}
\usepackage{epsfig}
\usepackage{latexsym}
\usepackage{paralist}
\usepackage{fancyhdr}
\usepackage{graphicx}
\numberwithin{equation}{section}
\usepackage[vcentermath]{youngtab}
\usepackage{young}
\usepackage{ytableau}
\usepackage{etex}
\usepackage{braket}
\usepackage{float}

\usepackage{pict2e}   
\usepackage{animate}  

\usepackage{multirow}
\usepackage{bigdelim}
\usepackage{fancybox}
\usepackage{cals}

\def\be{\begin{equation}}
\def\ee{\end{equation}}
\def\bea{\begin{eqnarray}}
\def\eea{\end{eqnarray}}

\renewcommand{\thefootnote}{\fnsymbol{footnote}}

\begin{document}

\hfuzz=100pt
\title{{\Large \bf{3d Self-Dualities} }}
\date{}
\author{ Keita Nii$^a$\footnote{nii@itp.unibe.ch}
}
\date{\today}

\maketitle

\thispagestyle{fancy}
\cfoot{}
\renewcommand{\headrulewidth}{0.0pt}

\vspace*{-1cm}
\begin{center}
$^{a}${{\it Albert Einstein Center for Fundamental Physics }}
\\{{\it Institute for Theoretical Physics
}}
\\ {{\it University of Bern}}  
\\{{\it  Sidlerstrasse 5, CH-3012 Bern, Switzerland}}

\end{center}

\begin{abstract}
We investigate self-dualities in three-dimensional $\mathcal{N}=2$ supersymmetric gauge theories. The electric and magnetic theories share the same gauge group. The examples include $SU(2N)$, $SO(7)$ and $SO(8)$ with various matter contents. The duality exchanges the role of the baryon and Coulomb branch operators in some examples. In other examples, the Coulomb branch operator becomes an elementary field on the dual side. These self-dualities in turn teach us a correct quantum structure of the Coulomb moduli space of vacua. Some dualities show symmetry enhancement. 
\end{abstract}

\renewcommand{\thefootnote}{\arabic{footnote}}
\setcounter{footnote}{0}

\newpage
\tableofcontents 
\clearpage

\newpage

\section{Introduction}
Duality is a very powerful tool of studying the low-energy dynamics of the strongly-coupled gauge theory. In supersymmetric gauge theories with four supercharges, this type of duality is called Seiberg duality \cite{Seiberg:1994bz, Seiberg:1994pq}. In general, the electric and magnetic (dual) theories have the different gauge groups and flow to the same infrared physics. 
In four dimensional spacetime, ``self-dualities'' where the electric and magnetic theories share the same gauge group were also found \cite{Csaki:1997cu, Karch:1997jp} in addition to the usual Seiberg dualities. 
It is important to study the self-dualities because these examples sometimes exhibit symmetry enhancement in the far-infrared limit and because the symmetry enhancement is in turn related to the existence of many sets of dualities \cite{Leigh:1996ds, Razamat:2016gzx, Razamat:2017hda, Razamat:2017wsk, Razamat:2018gbu}.

In this paper, we investigate the possibility of the self-dualities in the 3d $\mathcal{N}=2$ supersymmetric gauge theories with various gauge groups and various matter contents. In four dimensions, this was extensively studied in \cite{Csaki:1997cu, Karch:1997jp}. In three dimensions, this was recently investigated in\cite{Benvenuti:2018bav, Amariti:2018wht} for the $USp(2N)$ gauge theories with an anti-symmetric matter. The $SU(2)$ self-duality was studied in  \cite{Dimofte:2012pd}.
 We will mimic their approach and find similar self-dualities. In 4d, there are chiral anomalies and a particular $U(1)$ global symmetry is anomalous. The 4d duality does not respect this broken symmetry. Therefore, the naive dimensional reduction of the 4d self-dualities does not hold in 3d. However, we will find that the similar self-duality is indeed applicable. In 3d, there is an additional branch of the moduli space of vacua, which is called a Coulomb branch. The Coulomb brach is a flat direction of the scalar fields in the vector superfields. In \cite{Nii:2018bgf}, we studied the low-energy dynamics of the 3d $\mathcal{N}=2$ ``chiral'' $SU(N)$ theories with $F$ fundamental and $\bar{F}$ anti-fundamental matters. We there found that the ``chiral'' theory can have the rich structure of the quantum Coulomb branch, compared to the ``vector-like'' theory whose Coulomb branch is one-dimensional. The richness of the Coulomb branch can remedy seemingly-invalid dualities. In order to obtain the rich Coulomb branch, we have to slightly change the matter contents of the 4d self-duality. The resulting pair of the 3d theories has gauge anomalies if we assume that those self-dualities live in 4d. We will also discuss the connection between the 3d and 4d self-dualities. We will propose self-dualities for $SU(2N)$, $Spin(7)$ and $Spin(8)$ cases and these self-dualities in turn teach us the correct quantum structure of the Coulomb moduli space.


The rest of this paper is organized as follows. 
In Section 2, we will review the self-duality in the 3d $\mathcal{N}=2$ $SU(2)$ gauge theory with six fundamental matters, which would be the simplest case of the self-duality. 
In Section 3, we consider the self-duality in the 3d $\mathcal{N}=2$ $SU(4)$ gauge theory. 
In Section 4, we will consider the $SU(6)$ self-duality with a third-order anti-symmetric tensor.
In Section 5, we will consider the $SU(8)$ self-duality with two anti-symmetric tensors.
In Section 6, we will consider the $SU(2N)$ self-duality which generalizes the $SU(4)$ self-duality.
In Section 7, we will move on to the self-dualiies of the 3d $\mathcal{N}=2$ $Spin(N)$ gauge theories with vector and spinor matters.
In Section 8, we will summarize our findings and give possible future directions.

\section{$SU(2)$ self-duality}
For illustrating how the 3d self-dualities work, let us consider the 3d $\mathcal{N}=2$ $SU(2)$ gauge theory with six fundamental matters (six doublets). The low-energy dynamics of this theory was studied in \cite{Aharony:1997bx, deBoer:1997kr} and the self-duality was discussed in \cite{Dimofte:2012pd} (see also \cite{Benvenuti:2018bav, Amariti:2018wht}). The Higgs branch is described by a meson composite $M_{QQ}:=QQ$ while the Coulomb branch is parametrized by a single coordinate $Y$. When $Y$ obtains a non-zero vacuum expectation value, the gauge group is broken as $SU(2) \rightarrow U(1)$.
The matter contents and their quantum numbers are summarized in Table \ref{SU(2)F6}. The theory exhibits a manifest global $SU(6)$ symmetry. 

\begin{table}[H]\caption{3d $\mathcal{N}=2$ $SU(2)$ with $6 \, {\tiny \protect\yng(1)} $} 
\begin{center}
\scalebox{1}{
  \begin{tabular}{|c||c||c|c|c| } \hline
  &$SU(2)$&$SU(6)$&$U(1)$&$U(1)_R$  \\ \hline
$Q$&${\tiny \yng(1)}$ &${\tiny \yng(1)}$&1&$r$  \\ \hline
$M_{QQ}:=QQ$&1&${\tiny \yng(1,1)}$&2&$2r$ \\
$Y$&1&1&$-6$&$4-6r$ \\ \hline
  \end{tabular}}
  \end{center}\label{SU(2)F6}
\end{table}

We can regard this theory as the 3d $\mathcal{N}=2$ $SU(2)$ gauge theory with four fundamentals and two anti-fundamentals although the explicit $SU(6)$ flavor symmetry is invisible. The Coulomb branch is completely the same as the previous one while the Higgs branch operator is decomposed into the meson $M$, baryon $B$ and anti-baryon $\bar{B}$ operators in Table \ref{SU(2)F6expanded}. In the following subsections, we will review the $SU(2)$ self-dualities of Table \ref{SU(2)F6} or Table \ref{SU(2)F6expanded}.

\begin{table}[H]\caption{3d $\mathcal{N}=2$ $SU(2)$ with $4 \, {\tiny \protect\yng(1)} +2 \, {\tiny \overline{\protect\yng(1)}} $} 
\begin{center}
\scalebox{1}{
  \begin{tabular}{|c||c||c|c|c|c|c| } \hline
  &$SU(2)$&$SU(4)$&$SU(2)$&$U(1)$&$U(1)$&$U(1)_R$  \\ \hline
$Q$&${\tiny \yng(1)}$ &${\tiny \yng(1)}$&1&1&0&$r$  \\ 
$\tilde{Q}$&${\tiny \overline{\yng(1)}}={\tiny \yng(1)}$&1&${\tiny \yng(1)}$&0&1&$r$ \\ \hline
$M:=Q \tilde{Q}$&1&${\tiny \yng(1)}$&${\tiny \yng(1)}$&1&1&$2r$ \\
$B:=Q^2$&1&${\tiny \yng(1,1)}$&1&2&0&$2r$ \\
$\bar{B}:= \tilde{Q}^2$&1&1&1&0&2&$2r$ \\
$Y$&1&1&1&$-4$&$-2$&$4- 6r$ \\ \hline
  \end{tabular}}
  \end{center}\label{SU(2)F6expanded}
\end{table}

\subsection{$USp(2)=SU(2)$ self-dual}
First, we consider the $USp$ dual of the theory in Table \ref{SU(2)F6}. Since the $SU(2)$ group is a member of the symplectic groups, we can use the Aharony duality \cite{Aharony:1997gp} and obtain the $USp(2 \tilde{N})$ dual description. The dual gauge group is again $SU(2)$ in this case. The dual theory includes the meson and the Coulomb branch operator as elementary fields. Therefore, all the moduli coordinates are introduced as elementary fields on the dual side. 
\begin{table}[H]\caption{$USp(2)$ dual theory} 
\begin{center}
\scalebox{1}{
  \begin{tabular}{|c||c||c|c|c| } \hline
  &$SU(2)$&$SU(6)$&$U(1)$&$U(1)_R$  \\ \hline
$q$&${\tiny \yng(1)}$ &${\tiny \yng(1)}$&$-1$&$1-r$  \\ 
$M_{QQ}$&1&${\tiny \yng(1,1)}$&2&$2r$ \\
$Y$&1&1&$-6$&$4-6r$ \\ \hline
$\tilde{Y}$&1&1&$6$&$-2+6r$ \\ \hline
  \end{tabular}}
  \end{center}\label{USpSU(2)dual}
\end{table}
\noindent Table \ref{USpSU(2)dual} shows the matter contents and their quantum numbers of the $USp(2)$ dual theory. The dual theory has a tree-level superpotential
\begin{align}
W=M_{QQ}qq + Y \tilde{Y},
\end{align}
which is consistent with all the symmetries in Table \ref{USpSU(2)dual}. One can rewrite the $USp(2)$ dual in such a way that the global $SU(4) \times SU(2) \times U(1) \times U(1)$ symmetry is manifest (Table \ref{USpSU(2)dual2}). The superpotential is decomposed into 
\begin{align}
W= M q \tilde{q} +B q^2 +\bar{B} \tilde{q}^2 + Y \tilde{Y}.
\end{align}
In the following subsections, we will construct other self-dual descriptions where $M$ is only introduce as an elementary field or where $B, \bar{B}$ and $Y$ are introduced as elementary fields.
\begin{table}[H]\caption{$USp(2)$ dual theory in an $SU(4) \times SU(2)$ basis} 
\begin{center}
\scalebox{1}{
  \begin{tabular}{|c||c||c|c|c|c|c| } \hline
  &$SU(2)$&$SU(4)$&$SU(2)$&$U(1)$&$U(1)$&$U(1)_R$  \\ \hline
$q$&${\tiny \yng(1)}$ &${\tiny \overline{\yng(1)}}$&1&$-1$&0&$1-r$  \\ 
$\tilde{q}$&${\tiny \yng(1)}$&1&${\tiny \yng(1)}$&$0$&$-1$&$1-r$ \\ 
$M$&1&${\tiny \yng(1)}$&${\tiny \yng(1)}$&1&1&$2r$ \\
$B$&1&${\tiny \yng(1,1)}$&1&2&0&$2r$ \\
$ \bar{B}$&1&1&1&$0$&$2$&$2r$ \\ 
$Y$&1&1&1&$-4$&$-2$&$4 - 6 r$ \\  \hline
$\tilde{Y}$&1&1&1&$4$&$2$&$-2+6r$ \\ \hline
  \end{tabular}}
  \end{center}\label{USpSU(2)dual2}
\end{table}

\subsection{$SU(2)$ ``chiral'' self-dual}
Next, we consider the ``chiral'' self-dual description. By regarding the $SU(2)$ theory with $6 \, {\tiny \yng(1)}$ as an $SU(2)$ theory with four fundamentals and two anti-fundamentals (Table \ref{SU(2)F6expanded}), we can apply the ``chiral'' Seiberg duality \cite{Nii:2018bgf}. The dual gauge group again becomes $SU(2)$. The dual theory contains four fundamentals, two anti-fundamentals and a meson singlet $M$. The dual theory has a tree-level superpotential
\begin{align}
W=Mq \tilde{q}.
\end{align}
In this self-dual description, only the eight components of $M_{QQ}$, which are denoted by $M$, are introduced as elementary fields. 
The quantum numbers of the dual fields are summarized in Table \ref{SecondSU(2)dual}. The matching of the moduli operators is as follows. 
\begin{gather}
Q \tilde{Q} \sim  M, ~~~Q^2 \sim q^2 \nonumber \\
\tilde{Q}^2 \sim Y_{SU(2)},~~~Y \sim  \tilde{q}^2
\end{gather}
The role of the Coulomb branch and the anti-baryonic operator is exchanged. 

\begin{table}[H]\caption{``Chiral'' $SU(2)$ dual theory} 
\begin{center}
\scalebox{1}{
  \begin{tabular}{|c||c||c|c|c|c|c| } \hline
  &$SU(2)$&$SU(4)$&$SU(2)$&$U(1)$&$U(1)$&$U(1)_R$  \\ \hline
$q$&${\tiny \yng(1)}$ &${\tiny \overline{\yng(1)}}$&1&$1$&0&$r$  \\ 
$\tilde{q}$&${\tiny \overline{\yng(1)}}$&1&${\tiny \yng(1)}$&$-2$&$-1$&$2-3r$ \\ 
$M$&1&${\tiny \yng(1)}$&${\tiny \yng(1)}$&1&1&$2r$ \\ \hline
$B:=q^2$&1&${\tiny \yng(1,1)}$&1&2&0&$2r$ \\
$Y \sim \tilde{q}^2$&1&1&1&$-4$&$-2$&$4-6r$ \\ \hline
$\bar{B} \sim Y_{SU(2)}$&1&1&1&$0$&$2$&$2r$ \\ \hline
  \end{tabular}}
  \end{center}\label{SecondSU(2)dual}
\end{table}

\subsection{$SU(2)$ third self-dual}
We can construct the third self-dual description where the (anti-)baryonic operators and the Coulomb branch are introduced as elementary fields. The third dual description includes a tree-level superpotential
\begin{align}
W=B q^2 +Y \tilde{q}^2 +\bar{B} \tilde{Y}_{SU(2)},
\end{align}
which is consistent with all the symmetries in Table \ref{ThirdSU(2)dual}. The global $SU(6)$ symmetry is invisible and only the $SU(4) \times SU(2) \times U(1)$ subgroup is manifest. The global symmetry will be enhanced only in the infrared limit. The meson $M_{QQ}$ is decomposed into $B, \bar{B}$ and $M \sim q \tilde{q}$ in this self-dual description. The Coulomb branch $\tilde{Y}_{SU(2)}$ of the third self-dual is lifted by the superpotential.

\begin{table}[H]\caption{Third $SU(2)$ dual theory} 
\begin{center}
\scalebox{1}{
  \begin{tabular}{|c||c||c|c|c|c|c| } \hline
  &$SU(2)$&$SU(4)$&$SU(2)$&$U(1)$&$U(1)$&$U(1)_R$  \\ \hline
$q$&${\tiny \yng(1)}$ &${\tiny \yng(1)}$&1&$-1$&0&$1-r$  \\ 
$\tilde{q}$&${\tiny \yng(1)}$&1&${\tiny \yng(1)}$&$2$&$1$&$-1+3r$ \\ 
$B$&1&${\tiny \yng(1,1)}$&1&2&0&$2r$ \\
$\bar{B}$&1&1&1&$0$&$2$&$2r$ \\ 
$Y$&1&1&1&$-4$&$-2$&$4-6r$ \\ \hline
$M \sim q \tilde{q}$&1&${\tiny \yng(1)}$&${\tiny \yng(1)}$&1&1&$2r$ \\ \hline 
$\tilde{Y}_{SU(2)}$&1&1&1&$0$&$-2$&$2-2r$ \\ \hline
  \end{tabular}}
  \end{center}\label{ThirdSU(2)dual}
\end{table}

\section{$SU(4)$ self-duality \label{SU(4)}}
In this section, we consider the self-duality in the 3d $\mathcal{N}=2$ $SU(4)$ gauge theory with two anti-symmetric tensors and some (anti-)fundamental matters. We will find the two self-dual examples. The first one is the self-dual of the vector-like theory. The second one is chiral.  

\subsection{$SU(4)$ with $2\,\protect\Young[-0.5]{11}+3(\protect\Young[0]{1}+ \overline{\protect\Young[0]{1}})$}
The first example is the 3d $\mathcal{N}=2$ $SU(4)$ gauge theory with two anti-symmetric tensors and three flavors in the (anti-)fundamental representation. Since the anti-symmetric representation of $SU(4)$ is real, the theory can be regarded as a ``vector-like'' theory in a four-dimensional sense. The theory is also regarded as the $Spin(6)$ gauge theory and its Coulomb branch can be studied in the same manner as \cite{Csaki:2014cwa, Nii:2018wwj}. The Coulomb branch is now two-dimensional. The first coordinate corresponds to the breaking $SU(4) \rightarrow SU(2) \times U(1)_1 \times U(1)_2$ and is denoted as $Z$. The second one corresponds to the breaking $SU(4) \rightarrow SU(2) \times SU(2) \times U(1)$ and is denoted as $Y$. Since the theory is vector-like, these two operators are gauge-invariant. The Higgs branch is described by the following composite operators.
\begin{gather}
M_{0} :=Q \tilde{Q}, \quad M_2:=Q A^2\tilde{Q},  \quad T:=A^2  \nonumber \\
B:=Q^4,  \quad B_1:=AQ^2, \quad   \bar{B}_1:=A \tilde{Q}^2  
\end{gather}
Table \ref{SU(4)vectorelectric} summarizes the quantum numbers of the matter fields and the moduli coordinates. 

\begin{table}[H]\caption{3d $\mathcal{N}=2$ $SU(4)$ with $2\, {\tiny \protect\yng(1,1)} +3( \, {\tiny \protect\yng(1)}+ \,{\tiny \overline{\protect\yng(1)}})$} 
\begin{center}
\scalebox{0.9}{
  \begin{tabular}{|c||c||c|c|c|c|c|c|c| } \hline
  &$SU(4)$&$SU(2)$&$SU(3)$&$SU(3)$&$U(1)$&$U(1)$&$U(1)$&$U(1)_R$  \\ \hline
 $A$ &${\tiny \yng(1,1)}$&${\tiny \yng(1)}$&1&1&1&0&0&$r_A$ \\
 $Q$ & ${\tiny \yng(1)}$ &1& ${\tiny \yng(1)}$&1&0&1&0&$r$ \\
$\tilde{Q}$  &${\tiny \overline{\yng(1)}}$&1&1&${\tiny \yng(1)}$&0&$0$&1&$\bar{r}$ \\  \hline
$M_0:=Q\tilde{Q}$&1&1&${\tiny \yng(1)}$&${\tiny \yng(1)}$&0&1&1&$r+\bar{r}$ \\ 
$M_2:=QA^2\tilde{Q}$&1&1&${\tiny \yng(1)}$&${\tiny \yng(1)}$&2&1&1& $2r_A+r+\bar{r}$ \\
$T:=A^2$&1&${\tiny \yng(2)}$&1&1&2&0&0&$2r_A$ \\
$B_1:=AQ^2$&1&${\tiny \yng(1)}$&${\tiny \overline{\yng(1)}}$&1&1&2&0&$r_A +2r$ \\
$\bar{B}_1:=A \tilde{Q}^2$&1&${\tiny \yng(1)}$&1&${\tiny \overline{\yng(1)}}$&1&0&2&$r_A+2\bar{r}$  \\ \hline 
$Z:=Y_1Y_2Y_3$&1&1&1&1&$-4$&$-3$&$-3$&$4 -4r_A -3r -3 \bar{r}$ \\
$Y:= \sqrt{Y_1Y_2^2 Y_3}$&1&1&1&1&$-2$&$-3$& $-3$ & $4- 2 r_A -3r -3 \bar{r}$ \\ \hline
  \end{tabular}}
  \end{center}\label{SU(4)vectorelectric}
\end{table}

The dual description is again given by the 3d $\mathcal{N}=2$ $SU(4)$ gauge theory with two anti-symmetric tensors and three flavors in the (anti-)fundamnetal representation. The dual theory includes the gauge-singlet chiral superfields $M_0, M_2, B_1, \bar{B}_1, Y$ and $Z$ as elementary fields. These are identified with the moduli coordinates on the electric side. The magnetic description includes a tree-level superpotential
\begin{align}
W=M_0 q a^2 \tilde{q} +M_2  q \tilde{q} + B_1 aq^2 + \bar{B}_1 a \tilde{q}^2 +Y \tilde{Z} +Z \tilde{Y},  \label{WSU4}
\end{align}
where $\tilde{Z}, \tilde{Y}$ are the magnetic Coulomb branches corresponding to the breaking $SU(4) \rightarrow SU(2) \times U(1)_1 \times U(1)_2$ and $SU(4) \rightarrow SU(2) \times SU(2) \times U(1)$, respectively. The quantum numbers of the magnetic matter contents are summarized in Table \ref{SU(4)vectormagnetic}. The charge assignment of the magnetic matter is determined by the above superpotential. The matching of the chiral rings between the electric and magnetic theories is manifest from Table \ref{SU(4)vectormagnetic}. The non-trivial identification of the gauge invariant operator is $T:=A^2 \sim a^2$. 

Let us check the validity of this self-duality. First, the parity anomaly matching is satisfied. We can test the $USp(4)$ branch which can be realized by giving the expectation value to one of the anti-symmetric matters. Both the electric and magnetic theories flow to the $USp(4)$ theory with an anti-symmetric matter and six fundamentals. The $USp(4)$ self-duality was recently studied in \cite{Benvenuti:2018bav, Amariti:2018wht} and we can reproduce it. We can also connect this duality to the 4d self-duality of the 4d $\mathcal{N}=1$ $SU(4)$ theory \cite{Csaki:1997cu} with $2\, {\tiny \protect\yng(1,1)} +4( \, {\tiny \protect\yng(1)}+ \,{\tiny \overline{\protect\yng(1)}})$ via dimensional reduction and a real mass deformation \cite{Aharony:2013dha, Aharony:2013kma}. By putting the 4d self-dual pair on a circle, we obtain the self-duality with monopole superpotential. In order to get rid of the monopole effects, we introduce a positive real mass to one fundamental matter and a negative real mass to one anti-fundamental matter. The electric theory flows to Table \ref{SU(4)vectorelectric}. On the magnetic side, the real masses are introduced also for the gauge singlets. The massless components of the meson fields are decomposed into $M_0, M_2, Z$ and $Y$. The magnetic theory flows to Table \ref{SU(4)vectormagnetic}. The mechanism of the dynamical generation of the monopole superpotential in \eqref{WSU4} is unclear in this dimensional reduction process but it is consistent with all the symmetries. 

As another consistency check, we can compare the superconformal indices \cite{Bhattacharya:2008bja, Kim:2009wb, Imamura:2011su, Kapustin:2011jm} of the electric and magnetic theories. The both sides produce the same index

\scriptsize
\begin{align}
I&=1+x^{2/3} \left(\frac{1}{t^3 u^3 v^4}+9 t u+3 v^2\right) + x\left(6 t^2 v+6 u^2 v\right) \nonumber \\
&+x^{4/3} \left(\frac{1}{t^6 u^6 v^8} +\frac{9 t u+4 v^2}{t^3 u^3 v^4}+45 t^2 u^2+36 t u v^2+6 v^4\right)  +x^{5/3} \left(54 t^3 u v+18 t^2 v^3+\frac{6 \left(t^2+u^2\right)}{t^3 u^3 v^3}+54 t u^3 v+18 u^2 v^3\right) 
\nonumber \\
&+x^2 \left(\frac{1}{t^9 u^9 v^{12}}+\frac{9 t u+4 v^2}{t^6 u^6 v^8}+21 t^4 v^2+165 t^3 u^3+243 t^2 u^2 v^2+\frac{9 \left(5 t^2 u^2+4 t u v^2+v^4\right)}{t^3 u^3 v^4}+81 t u v^4+21 u^4 v^2+10 v^6-22\right) \nonumber \\
&+\cdots,
\end{align}
\normalsize

\noindent where $v$ is a fugacity for the first $U(1)$ symmetry of the anti-symmetric matters and $t,u$ are the fugacities for the remaining $U(1)$ symmetries. The r-charges are fixed to $r_A = r=\bar{r} =\frac{1}{3}$ for convenience. We computed the indices up to $O(x^{3})$ and found an exact agreement. The low-lying terms are easily identified with the moduli operators defined in Table \ref{SU(4)vectorelectric} and Table \ref{SU(4)vectormagnetic}. In the second term, $\frac{x^{2/3}}{t^3 u^3 v^4}$ corresponds to the Coulomb branch $Z$. In the fourth term, $\frac{4 x^{4/3}}{t^3 u^3 v^2}$ is interpreted as $Z T$ and $Y$.

\begin{table}[H]\caption{Dual of $SU(4)$ with $2\, {\tiny \protect\yng(1,1)} +3( \, {\tiny \protect\yng(1)}+ \,{\tiny \overline{\protect\yng(1)}})$} 
\begin{center}
\scalebox{0.9}{
  \begin{tabular}{|c||c||c|c|c|c|c|c|c| } \hline
  &$SU(4)$&$SU(2)$&$SU(3)$&$SU(3)$&$U(1)$&$U(1)$&$U(1)$&$U(1)_R$  \\ \hline
 $a$ &${\tiny \yng(1,1)}$&${\tiny \yng(1)}$&1&1&1&0&0&$r_A$ \\
 $q$ & ${\tiny \yng(1)}$ &1& ${\tiny \overline{\yng(1)}}$&1&$-1$&$-1$&0&$1-r_A -r$ \\
$\tilde{q}$  &${\tiny \overline{\yng(1)}}$&1&1&${\tiny \overline{\yng(1)}}$&$-1$&$0$&$-1$&$1-r_A-\bar{r}$ \\ 
$M_0$&1&1&${\tiny \yng(1)}$&${\tiny \yng(1)}$&0&1&$1$&$r+\bar{r}$ \\ 
$M_2$&1&1&${\tiny \yng(1)}$&${\tiny \yng(1)}$&2&1&1& $2r_A+r+\bar{r}$ \\
$B_1$&1&${\tiny \yng(1)}$&${\tiny \overline{\yng(1)}}$&1&1&2&0&$r_A +2r$ \\
$\bar{B}_1$&1&${\tiny \yng(1)}$&1&${\tiny \overline{\yng(1)}}$&1&0&2&$r_A+2\bar{r}$  \\ 
$Z$&1&1&1&1&$-4$&$-3$&$-3$&$4 -4r_A -3r -3 \bar{r}$ \\
$Y$&1&$1$&1&1&$-2$&$-3$& $-3$ & $4- 2 r_A -3r -3 \bar{r}$ \\ \hline
$T\sim a^2$&1&${\tiny \yng(2)}$&1&1&2&0&0&$2r_A$ \\ \hline 
$\tilde{Z}:= \tilde{Y}_1\tilde{Y}_2\tilde{Y}_3 $&1&1&1&1&$2$&$3$&$3$&$-2 +2r_A+3r+3 \bar{r}$ \\ 
$\tilde{Y}:= \sqrt{\tilde{Y}_1\tilde{Y}_2^2 \tilde{Y}_3} $&1&1&1&1&4&3&3&$-2+4r_A +3r+ 3 \bar{r}$ \\ \hline
  \end{tabular}}
  \end{center}\label{SU(4)vectormagnetic}
\end{table}

\subsection{$SU(4)$ with $2\,\protect\Young[-0.5]{11}+4\, \protect\Young[0]{1}+ 2 \,\overline{\protect\Young[0]{1}}$}

Next, we consider the self-duality in the 3d $\mathcal{N}=2$ $SU(4)$ gauge theory with two anti-symmetric matters, four fundamental matters and two anti-fundamental matters. The theory has no tree-level superpotential. The similar theory was studied in four-dimensions \cite{Csaki:1997cu}, where the theory was vector-like (four flavors) due to the gauge anomaly constraint. Of course, we can regard this theory as the $Spin(6)$ theory with two vectors, four spinors and two complex conjugate spinors. The ``chiral-ness'' of the theory will allow us to construct various self-dual description.

We first investigate the structure of the Coulomb moduli space of this theory. The bare Coulomb branch  leads to the gauge symmetry brealking
\begin{align}
SU(4) & \rightarrow SU(2) \times U(1)_1 \times U(1)_2  \\
{\tiny \yng(1,1)}  & \rightarrow {\tiny \yng(1)}_{1,0} +{\tiny \yng(1)}_{-1,0} + 1_{0,2}+1_{0,-2}  \\
{\tiny \yng(1)}  &\rightarrow  {\tiny \yng(1)}_{0,-1} +1_{1,1} +1_{-1,1}   \\
\overline{{\tiny \yng(1)}}  &\rightarrow  {\tiny \yng(1)}_{0,1} +1_{-1,-1} +1_{1,-1}
\end{align}
and we denote the corresponding operator as $Y^{bare}$. Along this Coulomb branch, the massive components are integrated out and the mixed Chern-Simons term is induced between the two $U(1)$ gauge groups
\begin{align}
k_{eff}^{U(1)_1U(2)}=2.
\end{align}
The mixed Chern-Simons term makes the bare Coulomb branch operator $Y^{bare}$ gauge non-invariant. The $U(1)_2$ charge of $Y^{bare}$ is $-2$. In order to construct the gauge invariant moduli coordinates, we can use the massless components from the anti-fundamental and anti-symmetric matters, $1_{0,2} \in {\tiny \yng(1,1)} $ or $ {\tiny \yng(1)}_{0,1} \in \overline{{\tiny \yng(1)}} $. The dressed (gauge-invariant) Coulomb branch operators become
\begin{align}
Y_A^{dressed} :=Y^{bare}A \\
Y_{\tilde{Q}}^{dressed} :=Y^{bare} \tilde{Q}^2,
\end{align}
where the flavor indices of the anti-quark chiral superfields are totally anti-symmetrized while $Y_A^{dressed}$ has a flavor index of the anti-symmetric matter. The Higgs branch is described by the following composite operators.
\begin{gather}
M_{0} :=Q \tilde{Q}, \quad M_2:=Q A^2\tilde{Q},  \quad T:=A^2  \nonumber \\
B:=Q^4,  \quad B_1:=AQ^2, \quad   \bar{B}_1:=A \tilde{Q}^2  
\end{gather}
Table \ref{SU(4)electric} shows the quantum numbers of the matter contents and the moduli coordinates.

\begin{table}[H]\caption{3d $\mathcal{N}=2$ $SU(4)$ with $2\, {\tiny \protect\yng(1,1)} +4 \, {\tiny \protect\yng(1)}+2 \,{\tiny \overline{\protect\yng(1)}}$} 
\begin{center}
\scalebox{0.8}{
  \begin{tabular}{|c||c||c|c|c|c|c|c|c| } \hline
  &$SU(4)$&$SU(2)$&$SU(4)$&$SU(2)$&$U(1)$&$U(1)$&$U(1)$&$U(1)_R$  \\ \hline
 $A$ &${\tiny \yng(1,1)}$&${\tiny \yng(1)}$&1&1&1&0&0&$r_A$ \\
 $Q$ & ${\tiny \yng(1)}$ &1& ${\tiny \yng(1)}$&1&0&1&0&$r$ \\
$\tilde{Q}$  &${\tiny \overline{\yng(1)}}$&1&1&${\tiny \yng(1)}$&0&$0$&1&$\bar{r}$ \\  \hline
$M_0:=Q\tilde{Q}$&1&1&${\tiny \yng(1)}$&${\tiny \yng(1)}$&0&1&1&$r+\bar{r}$ \\ 
$M_2:=QA^2\tilde{Q}$&1&1&${\tiny \yng(1)}$&${\tiny \yng(1)}$&2&1&1& $2r_A+r+\bar{r}$ \\
$T:=A^2$&1&${\tiny \yng(2)}$&1&1&2&0&0&$2r_A$ \\
$B:=Q^4$&1&1&1&1&0&4&0&$4r$ \\
$B_1:=AQ^2$&1&${\tiny \yng(1)}$&${\tiny \yng(1,1)}$&1&1&2&0&$r_A +2r$ \\
$\bar{B}_1:=A \tilde{Q}^2$&1&${\tiny \yng(1)}$&1&1&1&0&2&$r_A+2\bar{r}$  \\ \hline 
$Y^{bare}:=Y_1Y_2Y_3$&\footnotesize  $ U(1)_2$ charge: 2&1&1&1&$-4$&$-4$&$-2$&$4 -4 r_A -4r -2 \bar{r}$ \\
$Y_A^{dressed} :=Y^{bare}A$&1&${\tiny \yng(1)}$&1&1&$-3$&$-4$& $-2$ & $4- 3 r_A -4r -2 \bar{r}$ \\
$Y_{\tilde{Q}}^{dressed} :=Y^{bare} \tilde{Q}^2 $&1&1&1&1&$-4$&$-4$&0&$4 -4 r_A -4r $ \\ \hline
  \end{tabular}}
  \end{center}\label{SU(4)electric}
\end{table}

\subsubsection{First self-duality}
We start with the first dual description where the meson fields $M_0, M_2$, the baryon fields $B_1, \bar{B}_1$ and the dressed Coulomb branch $Y_A^{dressed}$ are introduced as elementary fields. The dual theory includes the tree-level superpotential
\begin{align}
W= M_0 qa^2 \tilde{q}^2 +M_2 q \tilde{q} + B_1 a q^2 + \bar{B}_1 a \tilde{q}^2 + Y_A^{dressed}\tilde{Y}^{dressed}_a.
\end{align}
The quantum numbers of the dual matter contents are summarized in Table \ref{SU(4)first}. The charge assignment is completely fixed by requiring that $a^2$ should be identified with $T:=A^2$ and from the above superpotential. 

The analysis of the Coulomb branch is the same as the electric theory but needs a different interpretation. The bare Coulomb branch $\tilde{Y}^{bare}$ corresponds to the breaking $SU(4) \rightarrow SU(2) \times U(1)_1 \times U(1)_2$ and requires the ``dressing'' procedure. The dressed operators are defined as
\begin{align}
\tilde{Y}_a^{dressed} := \tilde{Y}^{bare}a \\
\tilde{Y}_{\tilde{q}}^{dressed} := \tilde{Y}^{bare} \tilde{q}^2.
\end{align}
The Coulomb branch dressed by the anti-symmetric matter $a$ is lifted and excluded from the low-energy spectrum by the tree-level superpotential. The other operator $\tilde{Y}_{\tilde{q}}^{dressed}$ is identified with the baryon operator $B:=Q^4$. This is a bit surprising since $\tilde{Y}_{\tilde{q}}^{dressed}$ does not include the dual quark superfield at all. The dual baryon $q^4$ is identified with the dressed Coulomb branch $Y_{\tilde{Q}}^{dressed}$. The matching of the other operators is found in Table \ref{SU(4)first}. 

We can test various consistency conditions. The first self-dual description satisfies the parity anomaly matching. This 3d self-duality is connected to the 4d self-duality \cite{Csaki:1997cu} via dimensional reduction as in the previous subsection. In this case, we have to introduce the real masses to the $SU(2)$ subgroup of the $SU(4)$ (anti-quark) flavor symmetry. The electric theory flows to Table \ref{SU(4)electric}. On the magnetic side, the anti-baryon operator is decomposed into $\bar{B}_1$ and $Y_A^{dressed}$ and we recover the description in Table \ref{SU(4)first}. 

\begin{table}[H]\caption{First dual of $SU(4)$ with $2\, {\tiny \protect\yng(1,1)} +4 \, {\tiny \protect\yng(1)}+2 \,{\tiny \overline{\protect\yng(1)}}$} 
\begin{center}
\scalebox{0.8}{
  \begin{tabular}{|c||c||c|c|c|c|c|c|c| } \hline
  &$SU(4)$&$SU(2)$&$SU(4)$&$SU(2)$&$U(1)$&$U(1)$&$U(1)$&$U(1)_R$  \\  \hline
 $a$ &${\tiny \yng(1,1)}$&${\tiny \yng(1)}$&1&1&1&0&0&$r_A$ \\
 $q$ & ${\tiny \yng(1)}$ &1& ${\tiny  \overline{\yng(1)}}$&1&$-1$&$-1$&0&$1-r_A-r$ \\
$\tilde{q}$  &${\tiny \overline{\yng(1)}}$&1&1&${\tiny \yng(1)}$&$-1$&$0$&$-1$&$1-r_A-\bar{r}$ \\  
$M_0$&1&1&${\tiny \yng(1)}$&${\tiny \yng(1)}$&0&1&1&$r+\bar{r}$ \\ 
$M_2$&1&1&${\tiny \yng(1)}$&${\tiny \yng(1)}$&2&1&1& $2r_A+r+\bar{r}$ \\
$B_1$&1&${\tiny \yng(1)}$&${\tiny \yng(1,1)}$&1&1&2&0&$r_A +2r$ \\
$\bar{B}_1$&1&${\tiny \yng(1)}$&1&1&1&0&2&$r_A+2\bar{r}$  \\
$Y_A^{dressed} $&1&${\tiny \yng(1)}$&1&1&$-3$&$-4$& $-2$ & $4- 3 r_A -4r -2 \bar{r}$ \\ \hline
$q \tilde{q}$&1&1&${\tiny  \overline{\yng(1)}}$&${\tiny \yng(1)}$&$-2$&$-1$&$-1$&$2-2r_A -r - \bar{r}$ \\
$q a^2\tilde{q}$&1&1&${\tiny  \overline{\yng(1)}}$&${\tiny \yng(1)}$&0&$-1$&$-1$&$2-r-\bar{r}$ \\
$aq^2$&1&${\tiny \yng(1)}$&${\tiny \yng(1,1)}$&1&$-1$&$-2$&0&$2-r_A -2r$  \\
$a \tilde{q}^2$&1&${\tiny \yng(1)}$&1&1&$-1$&0&$-2$&$2-r_A -2 \bar{r}$ \\
$T \sim a^2$&1&${\tiny \yng(2)}$&1&1&2&0&0&$2r_A$ \\
$Y_{\tilde{Q}}^{dressed} \sim q^4$&1&1&1&1&$-4$&$-4$&0&$4- 4r_A -4r$ \\ \hline
$\tilde{Y}^{bare}:=\tilde{Y}_1\tilde{Y}_2\tilde{Y}_3$& \footnotesize  $ U(1)_2$ charge: 2&1&1&1&$2$&$4$&$2$&$-2 +2 r_A + 4 r + 2 \bar{r}$ \\
$\tilde{Y}_a^{dressed} := \tilde{Y}^{bare}a $&1&${\tiny \yng(1)}$&1&1&$3$&$4$& $2$ & $-2+ 3 r_A +4r +2 \bar{r}$ \\
$B\sim\tilde{Y}_{\tilde{q}}^{dressed} := \tilde{Y}^{bare} \tilde{q}^2 $&1&1&1&1&$0$&$4$&0&$4r $ \\ \hline
  \end{tabular}}
  \end{center}\label{SU(4)first}
\end{table}

\subsubsection{Second self-duality \label{second}}
Let us consider the second self-dual description. The dual description is given by the 3d $\mathcal{N}=2$ $SU(4)$ gauge theory with  two anti-symmetric matters, four fundamental matters, two anti-fundamental matters and two meson singlets $M_0, \, M_2$. The dual theory includes the tree-level superpotential
\begin{align}
W= M_0 q a^2 \tilde{q} +M_2 q \tilde{q}.
\end{align}
Table \ref{SU(4)SD2} shows the quantum numbers of the matter contents and the moduli coordinates. The charge assignment is determined by the above superpotential and by requiring the identification
\begin{align}
B:=Q^4 \sim q^4,  \quad T:=A^2 \sim a^2.
\end{align}
The Coulomb branch of the second dual description is defined in the same manner.
\begin{align}
\tilde{Y}_a^{dressed} := \tilde{Y}^{bare}a  \\
\tilde{Y}_{\tilde{q}}^{dressed} := \tilde{Y}^{bare} \tilde{q}^2
\end{align}
The operator matching is manifest from Table \ref{SU(4)SD2} and reads 
\begin{gather*}
B_1 :=  AQ^2  \sim aq^2,  \quad  \bar{B}_1 :=a \tilde{q}^2  \sim   \tilde{Y}_a^{dressed} \\
Y_A^{dressed} \sim a \tilde{q}^2,   \quad   Y_{\tilde{Q}}^{dressed}  \sim \tilde{Y}_{\tilde{q}}^{dressed}.
\end{gather*}

Let us study the validity of the second self-dual. The parity anomaly matching condition is satisfied. This 3d self-duality is also connected to the 4d self-duality \cite{Csaki:1997cu} as in the case of the first self-dual. 
By giving the expectation value to one of the anti-symmetric matters, we can reproduce the $USp(4)$ self-duality \cite{Benvenuti:2018bav, Amariti:2018wht}.

\begin{table}[H]\caption{Second dual of $SU(4)$ with $2\, {\tiny \protect\yng(1,1)} +4 \, {\tiny \protect\yng(1)}+2 \,{\tiny \overline{\protect\yng(1)}}$} 
\begin{center}
\scalebox{0.78}{
  \begin{tabular}{|c||c||c|c|c|c|c|c|c| } \hline
  &$SU(4)$&$SU(2)$&$SU(4)$&$SU(2)$&$U(1)$&$U(1)$&$U(1)$&$U(1)_R$  \\  \hline
 $a$ &${\tiny \yng(1,1)}$&${\tiny \yng(1)}$&1&1&1&0&0&$r_A$ \\
 $q$ & ${\tiny \yng(1)}$ &1& ${\tiny  \overline{\yng(1)}}$&1&$0$&$1$&0&$r$ \\
$\tilde{q}$  &${\tiny \overline{\yng(1)}}$&1&1&${\tiny \yng(1)}$&$-2$&$-2$&$-1$&$2-2r_A-r-\bar{r}$ \\  
$M_0$&1&1&${\tiny \yng(1)}$&${\tiny \yng(1)}$&0&1&1&$r+\bar{r}$ \\ 
$M_2$&1&1&${\tiny \yng(1)}$&${\tiny \yng(1)}$&2&1&1& $2r_A+r+\bar{r}$ \\ \hline
$q \tilde{q}$&1&1&${\tiny  \overline{\yng(1)}}$&${\tiny \yng(1)}$&$-2$&$-1$&$-1$&$2-2r_A -r - \bar{r}$ \\
$q a^2\tilde{q}$&1&1&${\tiny  \overline{\yng(1)}}$&${\tiny \yng(1)}$&0&$-1$&$-1$&$2-r-\bar{r}$ \\
$T \sim a^2$&1&${\tiny \yng(2)}$&1&1&2&0&0&$2r_A$ \\
$B \sim q^4$&1&1&1&1&$0$&$4$&0&$4r$ \\
$B_1 \sim aq^2$&1&${\tiny \yng(1)}$&${\tiny \yng(1,1)}$&1&$1$&$2$&0&$r_A +2r$  \\
$Y_A^{dressed} \sim a \tilde{q}^2$&1&${\tiny \yng(1)}$&1&1&$-3$&$-4$&$-2$&$4-3r_A -4r-2 \bar{r}$ \\ \hline
$\tilde{Y}^{bare}:=\tilde{Y}_1\tilde{Y}_2\tilde{Y}_3$& \footnotesize $U(1)_2$ charge: 2&1&1&1&$0$&$0$&$2$&$ 2 \bar{r}$ \\
$\bar{B_1} \sim \tilde{Y}_a^{dressed} := \tilde{Y}^{bare}a $&1&${\tiny \yng(1)}$&1&1&$1$&$0$& $2$ & $ r_A +2 \bar{r}$ \\
$Y_{\tilde{Q}}^{dressed}  \sim \tilde{Y}_{\tilde{q}}^{dressed} := \tilde{Y}^{bare} \tilde{q}^2 $&1&1&1&1&$-4$&$-4$&0&$4-4r_A-4r $ \\ \hline
  \end{tabular}}
  \end{center}\label{SU(4)SD2}
\end{table}

\subsubsection{Third self-duality}
Finally, we present the third self-dual description (Table \ref{SU(4)third}), where the baryon operators $B_1, \bar{B}_1$ and the Coulomb branch operator $Y_A^{dressed}$ are introduced as elementary fields. The dual theory includes a tree-level superpotential
\begin{align}
W=  B_1 a q^2 + \bar{B}_1 \tilde{Y}^{dressed}_a   + Y_A^{dressed}a \tilde{q}^2.
\end{align}
Notice that the anti-baryon $\bar{B}_1$ couples to the dual Coulomb branch operator $\tilde{Y}^{dressed}_a$ and the Coulomb branch field $ Y_A^{dressed}$ couples to the anti-baryon $a \tilde{q}^2$. Table \ref{SU(4)third} shows the quantum numbers of the dual matter fields and the moduli coordinates. The matching of the moduli fields is straightforward:
\begin{gather*}
M_0:= QQ \sim q \tilde{q},   \quad M_2:=Q A^2 \tilde{Q}  \sim q a^2 \tilde{q}, \quad T:=A^2  \sim a^2 \\
B:=Q^4  \sim  \tilde{Y}_{\tilde{q}}^{dressed}, \quad Y_{\tilde{Q}}^{dressed} \sim q^4
\end{gather*}

Let us check several consistencies. We can find that the dual description leads to the same parity anomalies as the electric side. This third self-duality is also connected to the 4d self-duality \cite{Csaki:1997cu} as in the case of the first and the second self-dualities. 
By giving the expectation value to one of the anti-symmetric matters, we can reproduce the $USp(4)$ self-duality studied in \cite{Benvenuti:2018bav, Amariti:2018wht}.

As a final consistency check, let us compute the superconformal indices \cite{Bhattacharya:2008bja, Kim:2009wb, Imamura:2011su, Kapustin:2011jm} of these four theories (one electric and three duals). We found that all the theories lead to the same indices and the result is

\scriptsize
\begin{align}
I &=1+x^{2/3} \left(8 t u+3 v^2\right)+x \left(\frac{2}{t^4 u^2 v^3}+12 t^2 v+2 u^2 v\right)+x^{4/3} \left(\frac{1}{t^4 v^4}+t^4+36 t^2 u^2+32 t u v^2+6 v^4\right) \nonumber \\
&+x^{5/3} \left(\frac{2 \left(8 t u+3 v^2\right)}{t^4 u^2 v^3}+96 t^3 u v+36 t^2 v^3+16 t u^3 v+6 u^2 v^3\right)  \nonumber \\
&+x^2 \left(\frac{3}{t^8 u^4 v^6}+8 t^5 u+78 t^4 v^2+\frac{4}{t^4 v^2}+120 t^3 u^3+\frac{8 u}{t^3 v^4}+190 t^2 u^2 v^2+\frac{18}{t^2 u^2 v^2}+72 t u v^4+3 u^4 v^2+10 v^6-24\right) +\cdots, 
\end{align}
\normalsize
\noindent where $v$ is the fugacity for the $U(1)$ symmetry of the anti-symmetric matter and $t,u$ are the fugacities for the $U(1)$ symmetries of the (anti-)fundamental matters. We set the r-charges as $r_A=r=\bar{r}=\frac{1}{3}$ for simplicity. We observed the agreement of the indices up to $O(x^3)$. The low-lying terms are easily identified with the moduli operators in Table \ref{SU(4)electric} and the symmetric products of them.

\begin{table}[H]\caption{Third dual of $SU(4)$ with $2\, {\tiny \protect\yng(1,1)} +4 \, {\tiny \protect\yng(1)}+2 \,{\tiny \overline{\protect\yng(1)}}$} 
\begin{center}
\scalebox{0.8}{
  \begin{tabular}{|c||c||c|c|c|c|c|c|c| } \hline
  &$SU(4)$&$SU(2)$&$SU(4)$&$SU(2)$&$U(1)$&$U(1)$&$U(1)$&$U(1)_R$  \\  \hline
 $a$ &${\tiny \yng(1,1)}$&${\tiny \yng(1)}$&1&1&1&0&0&$r_A$ \\
 $q$ & ${\tiny \yng(1)}$ &1& ${\tiny  \yng(1)}$&1&$-1$&$-1$&0&$1-r_A-r$ \\
$\tilde{q}$  &${\tiny \overline{\yng(1)}}$&1&1&${\tiny \yng(1)}$&$1$&$2$&1&$-1+r_A+2r+\bar{r}$ \\  
$B_1$&1&${\tiny \yng(1)}$&${\tiny \yng(1,1)}$&1&1&2&0&$r_A +2r$ \\
$\bar{B}_1$&1&${\tiny \yng(1)}$&1&1&1&0&2&$r_A+2\bar{r}$  \\
$Y_A^{dressed} $&1&${\tiny \yng(1)}$&1&1&$-3$&$-4$& $-2$ & $4- 3 r_A -4r -2 \bar{r}$ \\ \hline
$M_0 \sim q \tilde{q}$&1&1&${\tiny  \yng(1)}$&${\tiny \yng(1)}$&$0$&$1$&$1$&$r + \bar{r}$ \\
$M_2 \sim q a^2\tilde{q}$&1&1&${\tiny  \yng(1)}$&${\tiny \yng(1)}$&2&$1$&$1$&$2r_A +r+\bar{r}$ \\
$T \sim a^2$&1&${\tiny \yng(2)}$&1&1&2&0&0&$2r_A$ \\
$Y_{\tilde{Q}}^{dressed} \sim q^4$&1&1&1&1&$-4$&$-4$&0&$4- 4r_A -4r$ \\ 
$aq^2$&1&${\tiny \yng(1)}$&${\tiny \yng(1,1)}$&1&$-1$&$-2$&0&$2-r_A -2r$  \\
$a \tilde{q}^2$&1&${\tiny \yng(1)}$&1&1&$3$&4&$2$&$-2+3r_A +4r+2 \bar{r}$ \\ \hline
$\tilde{Y}^{bare}:=\tilde{Y}_1\tilde{Y}_2\tilde{Y}_3$& \footnotesize  $ U(1)_2$ charge: 2&1&1&1&$-2$&$0$&$-2$&$2 -2 r_A  - 2 \bar{r}$ \\
$\tilde{Y}_a^{dressed} := \tilde{Y}^{bare}a $&1&${\tiny \yng(1)}$&1&1&$-1$&$0$& $-2$ & $2- r_A -2 \bar{r}$ \\
$B\sim\tilde{Y}_{\tilde{q}}^{dressed} := \tilde{Y}^{bare} \tilde{q}^2 $&1&1&1&1&$0$&$4$&0&$4r $ \\ \hline
  \end{tabular}}
  \end{center}\label{SU(4)third}
\end{table}

\section{$SU(6)$ self-duality}
In this section, we will consider the self-duality in the $SU(6)$ gauge theory with a third-order (totally) anti-symmetric tensor. The four-dimensional version of this duality was studied in \cite{Csaki:1997cu}. We first discuss the self-duality and then derive the dualities between $SU(6)$ and $USp(4)$ gauge theories which can be obtained via a complex mass deformation in the subsequent subsections.

\subsection{$SU(6)$ with $\protect\Young[-0.8]{111}+6 \, \protect\Young[0]{1}+ 4 \,\overline{\protect\Young[0]{1}}$}
The electric theory is the 3d $\mathcal{N}=2$ $SU(6)$ gauge theory with a third-order anti-symmetric matter, six fundamental and four anti-fundamental matters. Notice that this theory is ``chiral'' and has a gauge anomaly in 4d while the 3d version is anomaly-free. The ``chiral-ness'' of the theory will make the Coulomb branch richer and hence the self-duality well works in 3d. In 4d, the $SU(6)$ gauge theory with a third-order anti-symmetric matter and six fundamental flavors was considered and the self-duality was proposed \cite{Csaki:1997cu}. In this case, the theory is ``vector-like'' and the corresponding 3d theory has a simple Coulomb branch structure and hence we could not find the self-duality.

As in the 4d case \cite{Csaki:1997cu}, the Higgs branch is described by the following eight composite operators
\begin{gather*}
M_0 := Q \tilde{Q},~~M_2:=QA^2 \tilde{Q},~~T_4:= A^4,~~B_0 :=Q^6\\
B_1 :=AQ^3,~~B_3:=A^3 Q^3,~~\bar{B}_1:=A \tilde{Q}^3,~~\bar{B}_3:=A^3 \tilde{Q}^3.
\end{gather*}
Table \ref{SU(6)electric} summarizes the quantum numbers of the matter contents and these moduli coordinates. Let us consider the Coulomb branch. When the bare Coulomb brach operator $Y_{SU(4)}^{bare}$ obtains the vacuum expectation value, the gauge group and the matter fields are decomposed into
\begin{align}
SU(6) & \rightarrow SU(4) \times U(1)_1 \times U(1)_2 \\
\mathbf{6} & \rightarrow \mathbf{4}_{0,-1} +\mathbf{1}_{1,2} + \mathbf{1}_{-1,2} \\
\overline{\mathbf{6}} & \rightarrow  \overline{\mathbf{4}}_{0,1} +\mathbf{1}_{-1,-2} + \mathbf{1}_{1,-2} \\
\mathbf{20} & \rightarrow \mathbf{6}_{1,0} +\mathbf{6}_{-1,0} +\overline{\mathbf{4}}_{0,-3}+\mathbf{4}_{0,3}.
\end{align}
The components with the $U(1)_1$ charge are massive and integrated out form the low-energy spectrum.
Since the theory is ``chiral'', the mixed Chern-Simons term between the two $U(1)$ subgroups is generated and then the bare Coulomb branch has a $U(1)_2$ charge. In order to construct the gauge invariant moduli operator, we can combine $Y_{SU(4)}^{bare}$ with $\overline{\mathbf{4}}_{0,1} \in \tilde{Q}$ and $\mathbf{4}_{0,3} \in A$. The dressed Coulomb branch operators becomes
\begin{align}
Y_{SU(4), \tilde{Q}}^{dressed}&:= Y_{SU(4)}^{bare} \tilde{Q}^4 \\
Y_{SU(4), \tilde{Q}A}^{dressed}&:= Y_{SU(4)}^{bare} \tilde{Q} A \\
Y_{SU(4), \tilde{Q}A^3}^{dressed}&:= Y_{SU(4)}^{bare} A^2 \tilde{Q} A.
\end{align}
Notice that $Y_{SU(4)}^{bare} A^2 \tilde{Q}A  \sim Y_{SU(4)}^{bare} (\overline{\mathbf{4}}_{0,-3} \mathbf{4}_{0,3}) \overline{\mathbf{4}}_{0,1} \mathbf{4}_{0,3}$ should be regarded as an independent operator since we cannot express this contribution in terms of $Y_{SU(4), \tilde{Q}A}^{dressed}$ and the Higgs branch operators. The dressed Coulomb branch can have the flavor indices and thier quantum numbers are summarized in Table \ref{SU(6)electric}.

\begin{table}[H]\caption{$SU(6)$ with $ {\tiny \protect\yng(1,1,1)} +6 \, {\tiny \protect\yng(1)}+4 \,{\tiny \overline{\protect\yng(1)}}$} 
\begin{center}
\scalebox{0.84}{
  \begin{tabular}{|c||c||c|c|c|c|c|c| } \hline
  &$SU(6)$&$SU(6)$&$SU(4)$&$U(1)$&$U(1)$&$U(1)$&$U(1)_R$  \\  \hline
 $A$ &${\tiny \yng(1,1,1)}$&1&1&1&0&0&$r_A$ \\
 $Q$ & ${\tiny \yng(1)}$ & ${\tiny  \yng(1)}$&1&$0$&$1$&0&$r$ \\
$\tilde{Q}$  &${\tiny \overline{\yng(1)}}$&1&${\tiny \yng(1)}$&$0$&$0$&1&$\bar{r}$ \\ \hline 
$M_0 := Q \tilde{Q}$&1&${\tiny \yng(1)}$ &${\tiny \yng(1)}$ &0&1&1&$r+\bar{r}$  \\
$M_2:=QA^2 \tilde{Q}$&1&${\tiny \yng(1)}$ &${\tiny \yng(1)}$ &2&1&1&$2r_A +r +\bar{r}$  \\
$T_4:= A^4$&1&1&1&4&0&0&$4r_A$  \\
$B_0 :=Q^6$&1&1&1&0&6&0&$6r$  \\
$B_1 :=AQ^3$&1&${\tiny \yng(1,1,1)}$ &1&1&3&0&$r_A+3r$   \\[5pt]  
$B_3:=A^3 Q^3$&1&${\tiny \yng(1,1,1)}$ &1&3&3&0&$3r_A+3r$  \\
$\bar{B}_1:=A \tilde{Q}^3$&1&1&${\tiny \overline{\yng(1)}}$&1&0&3&$r_A+ 3\bar{r}$ \\
$\bar{B}_3:=A^3 \tilde{Q}^3$&1&1&${\tiny \overline{\yng(1)}}$&3&0&3&$3r_A+3\bar{r}$  \\ \hline
$Y_{SU(4)}^{bare}$&\footnotesize $U(1)_2$ charge: $-4$&1&1&$-6$&$-6$&$-4$&$6-6r_A -6r -4\bar{r}$  \\
$Y_{SU(4), \tilde{Q}}^{dressed}:= Y_{SU(4)}^{bare} \tilde{Q}^4$&1&1&1&$-6$&$-6$&$0$& $6-6r_A -6r $ \\
$Y_{SU(4), \tilde{Q}A}^{dressed}:= Y_{SU(4)}^{bare} \tilde{Q} A$&1&1& ${\tiny \yng(1)}$ &$-5$&$-6$&$-3$&$6-5r_A -6r -3 \bar{r}$  \\
$Y_{SU(4), \tilde{Q}A^3}^{dressed}:= Y_{SU(4)}^{bare} A^2 \tilde{Q} A$&1&1& ${\tiny \yng(1)}$ &$-3$&$-6$&$-3$&$6-3r_A -6r -3 \bar{r}$  \\ \hline
  \end{tabular}}
  \end{center}\label{SU(6)electric}
\end{table}

The dual description is again the 3d $\mathcal{N}=2$ $SU(6)$ gauge theory with a third-order anti-symmetric matter, six fundamental and four anti-fundamental matters. In addition, the magnetic theory includes the mesonic fields $M_0, M_2$ as elementary fields with a tree-level superpotential
\begin{align}
W= M_0 q a^2\tilde{q} +M_2 q \tilde{q}.
\end{align}
The quantum numbers of the dual fields are summarized in Table \ref{SU(6)magnetic}. The charge assignment is determined by requiring the matching of the baryon operators constructed from the (dual) quarks $Q, q$ and from the superpotential above.

The Coulomb branch $\tilde{Y}_{SU(4)}^{bare}$ again corresponds to the breaking $SU(6) \rightarrow SU(4) \times U(1) \times U(1)$, which must be dressed by the dual matter fields. $ \tilde{Y}_{SU(4), \tilde{q}}^{dressed}:= \tilde{Y}_{SU(4)}^{bare} \tilde{q}^4$ is identified with $Y_{SU(4), \tilde{Q}}^{dressed}$ and the other Coulomb branch operators are mapped to the anti-baryon operators. The other operator matching is summarized in Table \ref{SU(6)magnetic}. The role of the anti-baryons and the Coulomb branch operators is exchanged. 

We can show several consistency checks of this self-duality. First, the parity anomaly matching condition is satisfied. In addition, this self-duality is connected to the 4d self-duality via dimensional reduction and via the real mass deformation \cite{Aharony:2013dha, Aharony:2013kma, Aharony:1997bx}. In 4d, the self-duality was constructed for the 4d $\mathcal{N}=1$ $SU(6)$ gauge theory with $ {\tiny \protect\yng(1,1,1)} +6 ( {\tiny \protect\yng(1)}+{\tiny \overline{\protect\yng(1)}})$. By putting this self-dual pair on a small circle, we get the self-duality with monopole superpotentials. In order to get rid of the monopole superpotential, we introduce the real masses to the $SU(2)$ subgroup of the second $SU(6)$ symmetry. The resulting dual pair exactly becomes the 3d duality discussed here. The matching of the moduli operators are highly changed in this flow.  

\begin{table}[H]\caption{Self-dual of $SU(6)$ with $ {\tiny \protect\yng(1,1,1)} +6 \, {\tiny \protect\yng(1)}+4 \,{\tiny \overline{\protect\yng(1)}}$} 
\begin{center}
\scalebox{0.8}{
  \begin{tabular}{|c||c||c|c|c|c|c|c| } \hline
  &$SU(6)$&$SU(6)$&$SU(4)$&$U(1)$&$U(1)$&$U(1)$&$U(1)_R$  \\  \hline
 $a$ &${\tiny \yng(1,1,1)}$&1&1&1&0&0&$r_A$ \\
 $q$ & ${\tiny \yng(1)}$ & ${\tiny \overline{ \yng(1)}}$&1&$0$&$1$&0&$r$ \\
$\tilde{q}$  &${\tiny \overline{\yng(1)}}$&1&${\tiny \overline{ \yng(1)} }$&$-2$&$-2$&$-1$&$2-2r_A -2r -\bar{r}$ \\ 
$M_0 $&1&${\tiny \yng(1)}$ &${\tiny \yng(1)}$ &0&1&1&$r+\bar{r}$  \\
$M_2$&1&${\tiny \yng(1)}$ &${\tiny \yng(1)}$ &2&1&1&$2r_A +r +\bar{r}$  \\ \hline
$T_4\sim  a^4$&1&1&1&4&0&0&$4r_A$  \\
$B_0 \sim q^6$&1&1&1&0&6&0&$6r$  \\
$B_1 \sim aq^3$&1&${\tiny \yng(1,1,1)}$ &1&1&3&0&$r_A+3r$   \\[5pt]  
$B_3 \sim a^3 q^3$&1&${\tiny \yng(1,1,1)}$ &1&3&3&0&$3r_A+3r$  \\
$Y_{SU(4), \tilde{Q}A}^{dressed} \sim a \tilde{q}^3$&1&1&${\tiny \yng(1)}$&$-5$&$-6$&$-3$&$6-5r_A -6r -3 \bar{r}$  \\
$Y_{SU(4), \tilde{Q}A^3}^{dressed}\sim A^3 \tilde{Q}^3$&1&1& ${\tiny \yng(1)}$ &$-3$&$-6$&$-3$&$6-3r_A -6r -3 \bar{r}$   \\ \hline
$\tilde{Y}_{SU(4)}^{bare}$&\footnotesize $U(1)_2$ charge: $-4$&1&1&$2$&$2$&$4$&$-2+2r_A +2r +4\bar{r}$  \\
$Y_{SU(4), \tilde{Q}}^{dressed} \sim \tilde{Y}_{SU(4), \tilde{q}}^{dressed}:= \tilde{Y}_{SU(4)}^{bare} \tilde{q}^4$&1&1&1&$-6$&$-6$&$0$& $6-6r_A -6r $ \\
$\bar{B}_1 \sim \tilde{Y}_{SU(4), \tilde{q}a}^{dressed}:= \tilde{Y}_{SU(4)}^{bare} \tilde{q} a$&1&1&${\tiny \overline{\yng(1)}}$&1&0&3&$r_A+ 3\bar{r}$  \\
$\bar{B}_3 \sim \tilde{Y}_{SU(4), \tilde{q}a^3}^{dressed}:= \tilde{Y}_{SU(4)}^{bare} a^2 \tilde{q} a$&1&1&${\tiny \overline{\yng(1)}}$&3&0&3&$3r_A+3\bar{r}$   \\ \hline
  \end{tabular}}
  \end{center}\label{SU(6)magnetic}
\end{table}

\subsection{$SU(6) \leftrightarrow USp(4)$ duality}
From the $SU(6)$ self-duality above, one can derive the dual description of the 3d $\mathcal{N}=2$ $SU(6)$ gauge theory with $ {\tiny \protect\yng(1,1,1)} +5 \, {\tiny \protect\yng(1)}+3 \,{\tiny \overline{\protect\yng(1)}}$. This is achieved by introducing a complex mass to one fundamental flavor. 
The electric theory flows to the $SU(6)$ theory with $ {\tiny \protect\yng(1,1,1)} +5 \, {\tiny \protect\yng(1)}+3 \,{\tiny \overline{\protect\yng(1)}}$. The analysis of the Higgs and Coulomb branch coordinates is very similar to the previous subsection. The results are summarized in Table \ref{SU(6)USp(4)}. Notice that $Y_{SU(4), \tilde{Q}}^{dressed} $ is not available since the number of anti-fundamentals is less than four. 

\begin{table}[H]\caption{$SU(6)$ with $ {\tiny \protect\yng(1,1,1)} +5 \, {\tiny \protect\yng(1)}+3 \,{\tiny \overline{\protect\yng(1)}}$} 
\begin{center}
\scalebox{0.85}{
  \begin{tabular}{|c||c||c|c|c|c|c|c| } \hline
  &$SU(6)$&$SU(5)$&$SU(3)$&$U(1)$&$U(1)$&$U(1)$&$U(1)_R$  \\  \hline
 $A$ &${\tiny \yng(1,1,1)}$&1&1&1&0&0&$r_A$ \\
 $Q$ & ${\tiny \yng(1)}$ & ${\tiny  \yng(1)}$&1&$0$&$1$&0&$r$ \\
$\tilde{Q}$  &${\tiny \overline{\yng(1)}}$&1&${\tiny \yng(1)}$&$0$&$0$&1&$\bar{r}$ \\ \hline 
$M_0 := Q \tilde{Q}$&1&${\tiny \yng(1)}$ &${\tiny \yng(1)}$ &0&1&1&$r+\bar{r}$  \\
$M_2:=QA^2 \tilde{Q}$&1&${\tiny \yng(1)}$ &${\tiny \yng(1)}$ &2&1&1&$2r_A +r +\bar{r}$  \\
$T_4:= A^4$&1&1&1&4&0&0&$4r_A$  \\
$B_1 :=AQ^3$&1&${\tiny \overline{\yng(1,1)}}$ &1&1&3&0&$r_A+3r$   \\[5pt]  
$B_3:=A^3 Q^3$&1&${\tiny \overline{\yng(1,1)}}$ &1&3&3&0&$3r_A+3r$  \\
$\bar{B}_1:=A \tilde{Q}^3$&1&1&1&1&0&3&$r_A+ 3\bar{r}$ \\
$\bar{B}_3:=A^3 \tilde{Q}^3$&1&1&1&3&0&3&$3r_A+3\bar{r}$  \\ \hline
$Y_{SU(4)}^{bare}$&\footnotesize $U(1)_2$ charge: $-4$&1&1&$-6$&$-5$&$-3$&$4-6r_A -5r -3\bar{r}$  \\
$Y_{SU(4), \tilde{Q}A}^{dressed}:= Y_{SU(4)}^{bare} \tilde{Q} A$&1&1& ${\tiny \yng(1)}$ &$-5$&$-5$&$-2$&$4-5r_A -5r -2 \bar{r}$  \\
$Y_{SU(4), \tilde{Q}A^3}^{dressed}:= Y_{SU(4)}^{bare} A^2 \tilde{Q} A$&1&1& ${\tiny \yng(1)}$ &$-3$&$-5$&$-2$&$4-3r_A -5r -2 \bar{r}$  \\ \hline
  \end{tabular}}
  \end{center}\label{SU(6)USp(4)}
\end{table}

On the magnetic side, the complex mass deformation corresponds to the Higgsing of the $SU(6)$ gauge symmetry. The $SU(6)$ magnetic gauge group is broken to $USp(4)$. The magnetic side flows to the 3d $\mathcal{N}=2$ $USp(4)$ gauge theory with an anti-symmetric matter and eight fundamentals. The theory also includes the gauge singlet fields $M_0$ and $M_2$. The global $SU(8)$ symmetry is explicitly broken by the tree-level superpotential
\begin{align}
W= M_0 q a \tilde{q} +M_2 q \tilde{q}.
\end{align}
The Higgs branch is described by five composite operators $a^2, q^2, aq^2, \tilde{q}^2$ and $a \tilde{q}^2$. The flat directions $q \tilde{q}$ and $qa^2 \tilde{q}$ are lifted by the above superpotential. The matching with the electric moduli fields are indicated in Table \ref{USp(4)dual}. The Coulomb branch was studied, for example, in \cite{Benvenuti:2018bav, Amariti:2018wht} (see also \cite{Amariti:2015vwa, Amariti:2015mva}) and now two-dimensional. When the Coulomb branch $Y_{USp(2)}$ gets an expectation value, the gauge group is broken $USp(2) \times U(1)$. The bare Coulomb branch is gauge invariant and described by $Y_{USp(2)}$ and $Y_{USp(2),a} :=Y_{USp(2)}a$. Notice that, under this breaking, the anti-symmetric matter is reduced to a massless singlet and $Y_{USp(2)}a$ should be regarded as an independent gauge invariant operator.  
These Coulomb branch operators are mapped to the anti-baryonic operators $\bar{B}_1$ and $\bar{B}_3$ respectively.

\begin{table}[H]\caption{$USp(4)$ magnetic dual of $SU(6)$ with $ {\tiny \protect\yng(1,1,1)} +5 \, {\tiny \protect\yng(1)}+3 \,{\tiny \overline{\protect\yng(1)}}$} 
\begin{center}
\scalebox{0.9}{
  \begin{tabular}{|c||c||c|c|c|c|c|c| } \hline
  &$USp(4)$&$SU(5)$&$SU(3)$&$U(1)$&$U(1)$&$U(1)$&$U(1)_R$  \\  \hline
 $a$ &${\tiny \yng(1,1)}$&1&1&2&0&0&$2r_A$ \\
 $q$ & ${\tiny \yng(1)}$ & ${\tiny \overline{ \yng(1)}}$&1&$\frac{1}{2}$&$\frac{3}{2}$&0&$\frac{1}{2}r_A +\frac{3}{2}r$ \\
$\tilde{q}$  &${\tiny \yng(1)}$&1&${\tiny \overline{ \yng(1)} }$&$-\frac{5}{2}$&$-\frac{5}{2}$&$-1$&$2-\frac{5}{2}r_A -\frac{5}{2}r -\bar{r}$ \\ 
$M_0 $&1&${\tiny \yng(1)}$ &${\tiny \yng(1)}$ &0&1&1&$r+\bar{r}$  \\
$M_2$&1&${\tiny \yng(1)}$ &${\tiny \yng(1)}$ &2&1&1&$2r_A +r +\bar{r}$  \\ \hline
$T_4 \sim a^2$&1&1&1&4&0&0&$4r_A$  \\
$B_1 \sim q^2$&1&${\tiny \overline{\yng(1,1)}}$&1&1&3&0&$r_A +3r$  \\
$B_3 \sim a q^2$&1&${\tiny \overline{\yng(1,1)}}$&1&3&3&0&$3r_A +3r$  \\
$Y_{SU(4), \tilde{Q}A}^{dressed} \sim \tilde{q}^2$&1&1&${\tiny \yng(1)}$&$-5$&$-5$&$-2$&$4 -5r_A -5r -2 \bar{r}$  \\
$Y_{SU(4), \tilde{Q}A^3}^{dressed} \sim a \tilde{q}^2$&1&1&${\tiny \yng(1)}$&$-3$&$-5$&$-2$&$4-3r_A -5r -2 \bar{r}$  \\ \hline
$\bar{B}_1 \sim Y_{USp(2)}$&1&1&1&$1$&0&3&$r_A +3 \bar{r}$  \\
$\bar{B}_3 \sim Y_{USp(2),a} :=Y_{USp(2)}a$&1&1&1&3&0&3&$3r_A +3 \bar{r}$   \\ \hline
  \end{tabular}}
  \end{center}\label{USp(4)dual}
\end{table}

\subsection{$USp(4) \leftrightarrow SU(6)$ duality}
By using the duality between the $SU(6)$ and $USp(4)$ gauge theories, we can take the $USp(4)$ theory as an electric description. The electric theory is the 3d $\mathcal{N}=2$ $USp(4)$ gauge theory with an anti-symmetric matter and eight fundamentals without a superpotential. Now, the $USp(4)$ theory has an explicit $SU(8)$ symmetry. However, we will keep only the $SU(5) \times SU(3)$ symmetry manifest in order to compare this theory with the magnetic theory. The quantum numbers of the matter fields are summarized in Table \ref{USp(4)/SU(6)electric}. The analysis of the moduli coordinates is identical to the previous case and the results are listed in Table \ref{USp(4)/SU(6)electric}.

\begin{table}[H]\caption{3d $\mathcal{N}=2$ $USp(4)$ with $ {\tiny \protect\yng(1,1)} +8 \, {\tiny \protect\yng(1)}$} 
\begin{center}
\scalebox{0.9}{
  \begin{tabular}{|c||c||c|c|c|c|c|c| } \hline
  &$USp(4)$&$SU(5)$&$SU(3)$&$U(1)$&$U(1)$&$U(1)$&$U(1)_R$  \\  \hline
 $A$ &${\tiny \yng(1,1)}$&1&1&1&0&0&$r_A$ \\
 $Q$ & ${\tiny \yng(1)}$ & ${\tiny  \yng(1)}$&1&$0$&$1$&0&$r$ \\
$\tilde{Q}$  &${\tiny \yng(1)}$&1&${\tiny \yng(1) }$&$0$&$0$&$1$&$\bar{r}$ \\   \hline
$M_m:=Q A^m\tilde{Q}~(m=0,1) $&1&${\tiny \yng(1)}$ &${\tiny \yng(1)}$ &$m$&1&1&$mr_A+r+\bar{r}$  \\
$B_m := A^mQ^2~(m=0,1) $&1&${\tiny \yng(1,1)}$&1&$m$&2&0&$mr_A +2r$  \\
$\bar{B}_m :=A^m \tilde{Q}^2~(m=0,1) $&1&1&${\tiny \overline{ \yng(1)}}$&$m$&0&2&$mr_A +2 \bar{r}$  \\ 
$T_2 :=A^2$&1&1&1&2&0&0&$2r_A$  \\ \hline
$Y_{USp(2)}$&1&1&1&$-2$&$-5$&$-3$&$6-2r_A -5r-3\bar{r}$  \\
$Y_{USp(2),A}:=Y_{USp(2)} A$&1&1&1&$-1$&$-5$&$-3$&$6-2r_A -5r-3\bar{r}$ \\ \hline
  \end{tabular}}
  \end{center}\label{USp(4)/SU(6)electric}
\end{table}

The magnetic side becomes the $SU(6)$ gauge theory with $ {\tiny \protect\yng(1,1,1)} +5 \, {\tiny \protect\yng(1)}+3 \,{\tiny \overline{\protect\yng(1)}}$ in addition to the gauge singlets $M_m~(m=0,1)$.
The tree-level superpotential becomes
\begin{align}
W= M_0 qa^2 \tilde{q} +M_1 q \tilde{q}.
\end{align}
The analysis of the moduli coordinates is identical to the previous case and the matching of the electric and magnetic moduli fields is indicated in Table \ref{USp(4)/SU(6)magnetic}. In this dual description, the global $SU(8)$ symmetry is not manifest and the symmetry will be enhanced to $SU(8)$ in the far-infrared limit.

\begin{table}[H]\caption{$SU(6)$ magnetic dual of $USp(4)$ with $ {\tiny \protect\yng(1,1)} +8 \, {\tiny \protect\yng(1)}$} 
\begin{center}
\scalebox{0.8}{
  \begin{tabular}{|c||c||c|c|c|c|c|c| } \hline
  &$SU(6)$&$SU(5)$&$SU(3)$&$U(1)$&$U(1)$&$U(1)$&$U(1)_R$  \\  \hline
 $a$ &${\tiny \yng(1,1,1)}$&1&1&$\frac{1}{2}$&0&0&$\frac{1}{2}r_A$ \\
 $q$ & ${\tiny \yng(1)}$ & ${\tiny \overline{ \yng(1)}}$&1&$-\frac{1}{6}$&$\frac{2}{3}$&0&$-\frac{1}{6}r_A +\frac{2}{3}r$ \\
$\tilde{q}$  &${\tiny \overline{\yng(1)}}$&1&${\tiny \overline{\yng(1)} }$&$-\frac{5}{6}$&$-\frac{5}{3}$&$-1$&$2 -\frac{5}{6}r_A -\frac{5}{3}r -\bar{r}$ \\   
$M_m~(m=0,1)$&1&${\tiny \yng(1)}$ &${\tiny \yng(1)}$ &$m$&1&1&$mr_A+r+\bar{r}$  \\  \hline
$B_m \sim a^{1+2m}q^3~(m=0,1)$&1&${\tiny \yng(1,1)}$&1&$m$&2&0&$mr_A +2r$  \\
$Y_{USp(2)} \sim a \tilde{q}^3$&1&1&1&$-2$&$-5$&$-3$&$6-2r_A -5r -3 \bar{r}$  \\ 
$Y_{USp(2),A} \sim a^3 \tilde{q}^3$&1&1&1&$-1$&$-5$&$-3$&$6-r_A -5r -3 \bar{r}$  \\ 
$T_2 \sim a^4$&1&1&1&2&0&0&$2r_A$  \\ \hline
$Y_{SU(4)}^{bare}$&\footnotesize $U(1)_2$ charge: $-4$&1&1&$\frac{1}{3}$&$\frac{5}{3}$&3&$-2 +\frac{1}{3}r_A +\frac{5}{3} r+3 \bar{r}$  \\
$\bar{B}_0 \sim Y_{SU(4),\tilde{q}a}^{dressed}:= Y_{SU(4)}^{bare} \tilde{q}a$&1&1&1&$0$&$0$&$2$&$2\bar{r}$  \\
$\bar{B}_1 \sim Y_{SU(4),\tilde{q}a^3}^{dressed}:= Y_{SU(4)}^{bare} \tilde{q}a^3$&1&1&1&$1$&$0$&$2$&$r_A +2\bar{r}$  \\ \hline
  \end{tabular}}
  \end{center}\label{USp(4)/SU(6)magnetic}
\end{table}

\section{$SU(8)$ self-duality}
The next example is the 3d $\mathcal{N}=2$ $SU(8)$ gauge theory with two anti-symmetric tensors and six anti-fundamental matters. The similar self-duality was studied in 4d \cite{Csaki:1997cu}. The Higgs branch is parametrized by three composite operators
\begin{gather}
\bar{B}_1 :=A \tilde{Q}^2,~~~\bar{B}_5 :=A^5 \tilde{Q}^2,~~~T_4:= A^4.
\end{gather}

The bare Coulomb branch $Y_{SU(6)}^{bare}$ corresponds to the gauge symmetry breaking
\begin{align}
SU(8) & \rightarrow SU(6) \times U(1)_1 \times U(1)_2  \\
{\tiny \overline{\yng(1)}} & \rightarrow {\tiny \overline{\yng(1)}}_{\, 0,1} +1_{-1,-3}+1_{1,-3} \\
{\tiny \yng(1,1)} & \rightarrow {\tiny \yng(1,1)}_{\, 0,-2} +{\tiny \yng(1)}_{\, 1,2}+{\tiny \yng(1)}_{\,-1,2}+1_{0,6}.
\end{align}
The components charged under the $U(1)_1$ subgroup are all massive and integrated out, which results in the mixed Chern-Simons term $k_{eff}^{U(1)_1, U(1)_2}= 6$. Therefore, the bare Coulomb branch $Y_{SU(6)}^{bare}$ is not gauge invariant and has the $U(1)_2$ charge. In order to construct the gauge invariant coordinates, we can use ${\tiny \overline{\yng(1)}}_{\, 0,1} \in {\tiny \overline{\yng(1)}}$ and $1_{0,6} \in {\tiny \yng(1,1)}$. The dressed Coulomb branch operators are defined as
\begin{align}
Y_{SU(6),\tilde{Q}}^{dressed}&:= Y_{SU(6)}^{bare} ({\tiny \overline{\yng(1)}}_{\, 0,1})^6 \sim Y_{SU(6)}^{bare} \tilde{Q}^6 \\
Y_{SU(6),A}^{dressed}&:= Y_{SU(6)}^{bare} 1_{0,6} \sim Y_{SU(6)}^{bare} A \\
Y_{SU(6),A^5}^{dressed}&:=Y_{SU(6)}^{bare} 1_{0,6}^2  \left({\tiny \yng(1,1)}_{\, 0,-2} \right)^3   \sim Y_{SU(6)}^{bare} A^5,
\end{align}
whose flavor indices are manifested in Table \ref{SU(8)electric}. Notice that $Y_{SU(6),A}^{dressed} \times T_4$ has ten independent components while $Y_{SU(6)}^{bare} 1_{0,6}^2  \left({\tiny \yng(1,1)}_{\, 0,-2} \right)^3$ has twelve components. The difference can be identified with an independent operator $Y_{SU(6),A^5}^{dressed}$.

\begin{table}[H]\caption{3d $\mathcal{N}=2$ $SU(8)$ gauge theory with $ 2\, {\tiny \protect\yng(1,1)} +6 \,{\tiny \overline{\protect\yng(1)}}$} 
\begin{center}
\scalebox{0.95}{
  \begin{tabular}{|c||c||c|c|c|c|c| } \hline
  &$SU(8)$&$SU(2)$&$SU(6)$&$U(1)$&$U(1)$&$U(1)_R$  \\ \hline
$A$&${\tiny \yng(1,1)}$ &${\tiny \yng(1)}$&1&$1$&0&$r_A$  \\ 
$\tilde{Q}$&${\tiny \overline{\yng(1)}}$&1&${\tiny \yng(1)}$&$0$&$1$&$\bar{r}$ \\  \hline 
$\bar{B}_1 :=A \tilde{Q}^2$&1&${\tiny \yng(1)}$&${\tiny \yng(1,1)}$&1&2&$r_A +2 \bar{r}$ \\
$\bar{B}_5 :=A^5 \tilde{Q}^2$&1&${\tiny \yng(1)}$&${\tiny \yng(1,1)}$&5&2&$5r_A +2\bar{r}$  \\ 
$T_4:= A^4$&1&${\tiny \yng(4)}$&1&4&0&$4r_A$  \\  \hline
$Y_{SU(6)}^{bare}$&\footnotesize $U(1)_2$ charge: $-6$&1&1&$-12$&$-6$&$4-12r_A -6\bar{r}$ \\
$Y_{SU(6),\tilde{Q}}^{dressed}:=Y_{SU(6)}^{bare} \tilde{Q}^6 $&1&1&1&$-12$&$0$&$4-12 r_A$  \\
$Y_{SU(6),A}^{dressed}:=Y_{SU(6)}^{bare} A$&1&${\tiny \yng(1)}$&1&$-11$&$-6$&$4 -11r_A -6 \bar{r}$ \\
$Y_{SU(6),A^5}^{dressed}:=Y_{SU(6)}^{bare} A^5$&1&${\tiny \yng(1)}$&1&$-7$&$-6$&$4 -7r_A -6 \bar{r}$ \\  \hline
  \end{tabular}}
  \end{center}\label{SU(8)electric}
\end{table}

Let us move on to the magnetic description which is again the 3d $\mathcal{N}=2$ $SU(8)$ gauge theory with two anti-symmetric tensors and four anti-fundamental matters. The elementary gauge singlet fields on the magnetic side are $\bar{B}_1,\bar{B}_5,Y_{SU(6),A}^{dressed}$ and $Y_{SU(6),A^5}^{dressed}$. The theory includes a tree-level superpotential 
\begin{align}
W= \bar{B}_1 a^5 \tilde{q}^2 +\bar{B}_5 a \tilde{q}^2 +Y_{SU(6),A}^{dressed}\tilde{Y}_{SU(6),a^5}^{dressed}+Y_{SU(6),A^5}^{dressed} \tilde{Y}_{SU(6),a}^{dressed}
\end{align}
which lifts the almost all the moduli fields of the magnetic side except for $T_4 \sim a^4$. We can easily check that the self-duality satisfies the parity anomaly matching condition. This self-duality can be related to the self-duality \cite{Csaki:1997cu} of the 4d $\mathcal{N}=1$ $SU(8)$ gauge theory with $ 2\, {\tiny \protect\yng(1,1)} +8 \,{\tiny \overline{\protect\yng(1)}}$ via dimensional reduction and the real mass deformation of the $SU(2)$ subgroup in the global $SU(8)$ symmetry. The massless components $\bar{B}_1^{78}$ and $\bar{B}_5^{78}$ of the singlet chiral superfields are identified with the two electric Coulomb branch operators, $Y_{SU(6),A}^{dressed}$ and $Y_{SU(6),A^5}^{dressed}$. Although the mechanism of the dynamical generation of the monopole potential is unclear, the above superpotential is consistent with all the symmetries. 

\begin{table}[H]\caption{Self-dual of $SU(8)$ with $ 2\, {\tiny \protect\yng(1,1)} +6 \,{\tiny \overline{\protect\yng(1)}}$} 
\begin{center}
\scalebox{0.9}{
  \begin{tabular}{|c||c||c|c|c|c|c| } \hline
  &$SU(8)$&$SU(2)$&$SU(6)$&$U(1)$&$U(1)$&$U(1)_R$  \\ \hline
$a$&${\tiny \yng(1,1)}$ &${\tiny \yng(1)}$&1&$1$&0&$r_A$  \\ 
$\tilde{q}$&${\tiny \overline{\yng(1)}}$&1&${\tiny \overline{\yng(1)}}$&$-3$&$-1$&$1-3r_A -\bar{r}$ \\   
$\bar{B}_1 $&1&${\tiny \yng(1)}$&${\tiny \yng(1,1)}$&1&2&$r_A +2 \bar{r}$ \\
$\bar{B}_5 $&1&${\tiny \yng(1)}$&${\tiny \yng(1,1)}$&5&2&$5r_A +2\bar{r}$  \\
$Y_{SU(6),A}^{dressed}$&1&${\tiny \yng(1)}$&1&$-11$&$-6$&$4 -11r_A -6 \bar{r}$ \\
$Y_{SU(6),A^5}^{dressed}$&1&${\tiny \yng(1)}$&1&$-7$&$-6$&$4 -7r_A -6 \bar{r}$ \\  \hline
$T_4\sim a^4$&1&${\tiny \yng(4)}$&1&4&0&$4r_A$  \\  \hline
$\tilde{Y}_{SU(6)}^{bare}$&\footnotesize $U(1)_2$ charge: $-6$&1&1&$6$&$6$&$-2+6r_A +6\bar{r}$ \\
$Y_{SU(6),\tilde{Q}}^{dressed} \sim \tilde{Y}_{SU(6),\tilde{q}}^{dressed}:=\tilde{Y}_{SU(6)}^{bare} \tilde{q}^6 $&1&1&1&$-12$&$0$&$4-12 r_A$  \\
$\tilde{Y}_{SU(6),a}^{dressed}:= \tilde{Y}_{SU(6)}^{bare} a$&1&${\tiny \yng(1)}$&1&$7$&$6$&$-2 +7r_A +6 \bar{r}$ \\
$\tilde{Y}_{SU(6),a^5}^{dressed}:= \tilde{Y}_{SU(6)}^{bare} a^5$&1&${\tiny \yng(1)}$&1&$11$&$6$&$-2 +11r_A +6 \bar{r}$ \\  \hline
  \end{tabular}}
  \end{center}\label{SU(8)magnetic}
\end{table}

As a consistency check of the self-duality, we can derive the self-duality of the 3d $\mathcal{N}=2$ $USp(8)$ theory with one anti-symmetric matter and six fundamentals \cite{Benvenuti:2018bav, Amariti:2018wht}. This flow can be achieved by introducing the vev to one of the anti-symmetric matters as follows.
\begin{align}
\braket{A_2} = v \begin{pmatrix}
 i \sigma_2 &0&0&0 \\
0&i \sigma_2 &0&0\\
0&0&i \sigma_2 &0\\
0&0&0& i\sigma_2
\end{pmatrix}
\end{align}
The electric $SU(8)$ theory is Higgsed to the $USp(8)$ theory with ${\tiny \yng(1,1)}+6 \, {\tiny \yng(1)}$. On the magnetic side, the vev for $A_2$ is mapped to the vev of $a_2$ with the same form 
\begin{align}
\braket{a_2} = v' \begin{pmatrix}
 i \sigma_2 &0&0&0 \\
0&i \sigma_2 &0&0\\
0&0&i \sigma_2 &0\\
0&0&0& i\sigma_2
\end{pmatrix}.
\end{align}
The magnetic side becomes the $USp(8)$ theory with ${\tiny \yng(1,1)}+6 \, {\tiny \yng(1)}$. The gauge singlet chiral superfields are decomposed into the singlets introduced in \cite{Benvenuti:2018bav, Amariti:2018wht} and the same superpotential is reproduced.

\section{$SU(2N)$ self-duality}
In this section, we will generalize the self-duality of the $SU(4)$ gauge theory with $2\, {\tiny \yng(1,1)}+4 \,{\tiny \yng(1)}+2\, {\tiny \overline{\yng(1)}}$ which was discussed in Section \ref{SU(4)}. The electric theory is the 3d $\mathcal{N}=2$ $SU(2N)$ gauge theory with an anti-symmetric flavor, four fundamental and two anti-fundamental matters. The matter contents and their quantum numbers are summarized in Table \ref{SU(2N)electric}. The similar 4d theory and its self-dual were studied in \cite{Csaki:1997cu}. The Higgs branch is the same as the 4d case with a small modification and described by 
\begin{gather*}
M_k:= \tilde{Q}(A \bar{A})^kQ,~~H_m:= \bar{A} (A \bar{A})^m Q^2,~~,\bar{H}_m:= A (\bar{A}A)^m \tilde{Q}^2 \\
B_N:=A^N,~~B_{N-1}:= A^{N-1} Q^2,~~B_{N-2}:=A^{N-2}Q^4 \\
\bar{B}_N:= \bar{A}^N,~~\bar{B}_{N-1}:= \bar{A}^{N-1} \tilde{Q}^2,~~\bar{B}_{N-1}:= \bar{A}^{N-1} \tilde{Q}^2,~~T_n:=(A \bar{A})^n,
\end{gather*}
where $k=0,\cdots,N-1$, $m=0,\cdots,N-2$ and $n=1,\cdots,N-1$.
Let us consider the Coulomb branch which was absent in 4d. The bare Coulomb branch operator $Y_{SU(2N-2)}^{bare}$ leads to the gauge symmetry breaking
\begin{align*}
SU(2N) & \rightarrow SU(2N-2) \times U(1)_1 \times U(1)_2  \\
{\tiny \yng(1)} & \rightarrow {\tiny \yng(1)}_{\, 0,-1} +1_{1,N-1}+1_{-1,N-1} \\
{\tiny \overline{\yng(1)}} & \rightarrow {\tiny \overline{\yng(1)}}_{\, 0,1} +1_{-1,-(N-1)}+1_{1,-(N-1)} \\
{\tiny \yng(1,1)} & \rightarrow {\tiny \yng(1,1)}_{\, 0,-2} +{\tiny \yng(1)}_{\, 1,N-2} +{\tiny \yng(1)}_{\, -1, N-2} +1_{0,2(N-1)} \\
{\tiny \overline{\yng(1,1)}} & \rightarrow {\tiny  \overline{\yng(1,1)}}_{\, 0,2} +{\tiny \overline{\yng(1)}}_{\, -1,-(N-2)} +{\tiny \overline{\yng(1)}}_{\, 1, -(N-2)} +1_{0,-2(N-1)},
\end{align*}
where the Coulomb branch corresponds to the $U(1)_1$ subgroup. The components with a non-zero $U(1)_1$ charge are massive and integrated out from the low-energy spectrum. Since the theory is ``chiral'', the mixed Chern-Simons term is generated between the $U(1)_1$ and $U(1)_2$ subgroups. The mixed Chern-Simons term makes the bare Coulomb branch operator charged under the $U(1)_2$ gauge group. In order to construct the gauge invariant moduli coordinate from the bare Coulomb branch, we can use various massless components of the matter fields and define the Higgs-Coulomb composite operators. The dressed Coulomb branch operators are defined in Table \ref{SU(2N)electric}. They have no flavor index and are only charged under the global $U(1)$ symmetries.

\begin{table}[H]\caption{3d $\mathcal{N}=2$ $SU(2N)$ with $ {\tiny \protect\yng(1,1)}+{\tiny \overline{\protect\yng(1,1)}}  +4 \, {\tiny \protect\yng(1)}+ 2\,{\tiny \overline{\protect\yng(1)}}$} 
\begin{center}
\scalebox{0.52}{
  \begin{tabular}{|c||c||c|c|c|c|c|c|c| } \hline
  &$SU(2N)$&$SU(4)$&$SU(2)$&$U(1)$&$U(1)$&$U(1)$&$U(1)$&$U(1)_R$  \\ \hline
 $A$ &${\tiny \yng(1,1)}$&1&1&1&0&0&0&$r_A$ \\
 $\bar{A}$ &${\tiny \overline{ \yng(1,1)}}$&1&1&0&1&0&0&$\bar{r}_A$ \\
 $Q$ & ${\tiny \yng(1)}$ &${\tiny \yng(1)}$& 1&0&0&1&0&$r$ \\
$\tilde{Q}$  &${\tiny \overline{\yng(1)}}$&1&${\tiny \yng(1)}$&0&$0$&0&1&$\bar{r}$ \\  \hline
$M_k:= \tilde{Q}(A \bar{A})^kQ$&1&${\tiny \yng(1)}$&${\tiny \yng(1)}$&$k$&$k$&1&1&$kr_A +k \bar{r}_A+r+\bar{r}$ \\ 
$H_m:= \bar{A} (A \bar{A})^m Q^2$&1&${\tiny \yng(1,1)}$&1&$m$&$m+1$&$2$&0&$mr_A +(m+1) \bar{r}_{A}+2r$  \\
$\bar{H}_m:= A (\bar{A}A)^m \tilde{Q}^2$&1&1&1&$m+1$&$m$&0&2&$(m+1)r_A +m \bar{r}_A +2 \bar{r}$  \\
$B_N:=A^N$ &1&1&1&$N$&0&0&0&$Nr_A$ \\
$B_{N-1}:= A^{N-1} Q^2$&1&${\tiny \yng(1,1)}$&1&$N-1$&0&2&0&$(N-1)r_A +2r$  \\
$B_{N-2}:=A^{N-2}Q^4$&1&1&1&$N-2$&0&4&0&$(N-2)r_A +4r$  \\
$\bar{B}_N:= \bar{A}^N$&1&1&1&0&$N$&0&0&$N \bar{r}_A$  \\
$\bar{B}_{N-1}:= \bar{A}^{N-1} \tilde{Q}^2$&1&1&1&0&$N-1$&0&2&$(N-1)\bar{r}_A +2 \bar{r}$  \\
$T_n:=(A \bar{A})^n$&1&1&1&$n$&$n$&0&0&$nr_A +n\bar{r}_A$  \\ \hline
$Y_{SU(2N-2)}^{bare}$&\footnotesize $U(1)_2$ charge: $-2(N-1)$&1&1&$-(2N-2)$&$-(2N-2)$&$-4$&$-2$&$4-2(N-1)r_A -2(N-1)\bar{r}_A -4r -2 \bar{r}$  \\
$Y_{SU(2N-2), m}^{dressed}:= Y_{SU(2N-2)}^{bare} A(A \bar{A})^m$&1&1&1&$m+3-2N$&$m+2-2N$&$-4$&$-2$&$4+(m+3-2N)r_A +(m+2-2N) \bar{r}_A -4r -2 \bar{r}$ \\ 
$Y_{SU(2N-2), \bar{A}}^{dressed}:= Y_{SU(2N-2)}^{bare} \bar{A}^{N-1}$&1&1&1&$2-2N$&$1-N$&$-4$&$-2$&$4+(2-2N)r_A +(1-N)\bar{r}_A -4r -2 \bar{r}$ \\ 
$Y_{SU(2N-2), \tilde{Q}}^{dressed}:= Y_{SU(2N-2)}^{bare} \bar{A}^{N-2}\tilde{Q}^2 $&1&1&1&$2-2N$&$-N$&$-4$&0&$4 +(2-2N)r_A -N \bar{r}_A -4r$ \\   \hline
  \end{tabular}}
  \end{center}\label{SU(2N)electric}
\end{table}

The magnetic description is again given by the 3d $\mathcal{N}=2$ $SU(2N)$ gauge theory with an anti-symmetric flavor, four fundamental and two anti-fundamental matters. By following the analysis  \cite{Csaki:1997cu} of the 4d $\mathcal{N}=1$ $SU(2N)$ self-duality, the magnetic theory here includes only the meson operators $M_k$ as elementary fields. The magnetic theory contains a tree-level superpotential
\begin{align}
W= \sum_{k=0}^{N-1} M_k   \tilde{q} (a \bar{a})^{N-1-k} q,
\end{align}
which is consistent with all the symmetries in Table \ref{SU(2N)magnetic}. The analysis of the magnetic theory is the same as the electric one and summarized in Table \ref{SU(2N)magnetic}. 

\begin{table}[H]\caption{Self-dual of $SU(2N)$ with $ {\tiny \protect\yng(1,1)}+{\tiny \overline{\protect\yng(1,1)}}  +4 \, {\tiny \protect\yng(1)}+ 2\,{\tiny \overline{\protect\yng(1)}}$} 
\begin{center}
\scalebox{0.48}{
  \begin{tabular}{|c||c||c|c|c|c|c|c|c| } \hline
  &$SU(2N)$&$SU(4)$&$SU(2)$&$U(1)$&$U(1)$&$U(1)$&$U(1)$&$U(1)_R$  \\ \hline
 $a$ &${\tiny \yng(1,1)}$&1&1&1&0&0&0&$r_A$ \\
 $\bar{a}$ &${\tiny \overline{ \yng(1,1)}}$&1&1&0&1&0&0&$\bar{r}_A$ \\
 $q$ & ${\tiny \yng(1)}$ &${\tiny \overline{\yng(1)}}$& 1&0&0&1&0&$r$ \\
$\tilde{q}$  &${\tiny \overline{\yng(1)}}$&1&${\tiny \yng(1)}$&$-(N-1)$&$-(N-1)$&$-2$&$-1$&$2- (N-1)r_A -(N-1) \bar{r}_A -2r -\bar{r}$ \\ 
$M_k$&1&${\tiny \yng(1)}$&${\tiny \yng(1)}$&$k$&$k$&1&1&$kr_A+k\bar{r}_A +r+\bar{r}$ \\  \hline
$ H_m \sim \bar{a} (a \bar{a})^m q^2$&1&${\tiny \yng(1,1)}$&1&$m$&$m+1$&$2$&0&$mr_A +(m+1) \bar{r}_{A}+2r$  \\
$Y_{SU(2N-2), m}^{dressed} \sim a (\bar{a}a)^m \tilde{q}^2$&1&1&1&$m+3-2N$&$m+2-2N$&$-4$&$-2$&$4+(m+3-2N)r_A +(m+2-2N) \bar{r}_A -4r -2 \bar{r}$   \\
$B_N \sim a^N$ &1&1&1&$N$&0&0&0&$Nr_A$ \\
$B_{N-1} \sim a^{N-1} q^2$&1&${\tiny \yng(1,1)}$&1&$N-1$&0&2&0&$(N-1)r_A +2r$  \\
$B_{N-2} \sim a^{N-2}q^4$&1&1&1&$N-2$&0&4&0&$(N-2)r_A +4r$  \\
$\bar{B}_N \sim \bar{a}^N$&1&1&1&$0$&$N$&$0$&$0$&$N \bar{r}_A$  \\
$Y_{SU(2N-2), \bar{A}}^{dressed} \sim  \bar{a}^{N-1} \tilde{q}^2$&1&1&1&$2-2N$&$1-N$&$-4$&$-2$&$4+(2-2N)r_A +(1-N)\bar{r}_A -4r -2 \bar{r}$ \\
$T_n \sim (a \bar{a})^n$&1&1&1&$n$&$n$&0&0&$nr_A + n\bar{r}_A$  \\ \hline
$  \tilde{Y}_{SU(2N-2)}^{bare}$&\footnotesize $U(1)_2$ charge: $-2(N-1)$&1&1&$0$&$0$&$0$&$2$&$2 \bar{r}$  \\
$\bar{H}_m \sim \tilde{Y}_{SU(2N-2), m}^{dressed}:= \tilde{Y}_{SU(2N-2)}^{bare} a(a \bar{a})^m$&1&1&1&$m+1$&$m$&0&2&$(m+1)r_A +m \bar{r}_A +2 \bar{r}$ \\ 
$\bar{B}_{N-1} \sim  \tilde{Y}_{SU(2N-2), \bar{a}}^{dressed}:= \tilde{Y}_{SU(2N-2)}^{bare} \bar{a}^{N-1}$&1&1&1&0&$N-1$&0&2&$(N-1)\bar{r}_A +2 \bar{r}$ \\ 
$ Y_{SU(2N-2), \tilde{Q}}^{dressed} \sim \tilde{Y}_{SU(2N-2), \tilde{q}}^{dressed}:= Y_{SU(2N-2)}^{bare} \bar{a}^{N-2}\tilde{q}^2 $&1&1&1&$2-2N$&$-N$&$-4$&0&$4 +(2-2N)r_A -N \bar{r}_A -4r$ \\   \hline
  \end{tabular}}
  \end{center}\label{SU(2N)magnetic}
\end{table}

The matching of the moduli coordinates is described in Table \ref{SU(2N)magnetic}. The role of the Coulomb and Higgs branch (constructed from the anti-quarks) operators is exchanged under the duality. The parity anomaly matching is satisfied on both sides. As a further consistency check, when $A$ and $\bar{A}$ obtain the vacuum expectation values, the both theories flow to the $SU(2)^N$ gauge theory, where each gauge group contains six fundamentals. As known in \cite{Dimofte:2012pd}, the $SU(2)$ theory with six fundamentals has the self-dual description. For $N=2$, the magnetic self-dual corresponds to the second self-dual description in Section \ref{second}. We can derive this self-duality from the 4d self-duality as in Section \ref{SU(4)} by putting the 4d self-dual pair on a circle and by introducing the real masses to two anti-fundamental matters.

\section{$Spin(N)$ self-duality}
In this section, we study the self-dualities in the 3d $\mathcal{N}=2$ $Spin(N)$ gauge theory with vector matters and spinor matters. The corresponding 4d self-dualities were investigated in \cite{Csaki:1997cu, Karch:1997jp}. By following the same spirit as \cite{Csaki:1997cu, Karch:1997jp}, we will try to construct the $Spin(N)$ self-dualities. We could find the 3d self-dualities only for $Spin(7)$ and $Spin(8)$ cases. The self-dual examples in 3d contain one spinor matter fewer than the 4d cases. In 4d, the self-dualities were constructed also for $Spin(N \ge 9)$ cases. However, in 3d, those one-spinor-less theories show s-confinement \cite{Nii:2018wwj} and the self-duality seems not available. The analysis of the $Spin(N \ge 9)$ self-dualities would be left as a future direction and we here focus on the $Spin(7)$ and $Spin(8)$ cases. 

\subsection{$Spin(7)$ self-duality}
Let us first consider the self-duality in the 3d $\mathcal{N}=2$ $Spin(7)$ gauge theory with three vectors and three spinors. The Coulomb branch of the $Spin(7)$ theory was studied in \cite{Nii:2018tnd} (see also \cite{Aharony:2011ci, Aharony:2013kma}). The Coulomb branch is now two-dimensional and parametrized by the two operators $Y, Z$. When the operator $Y$ obtains a non-zero vacuum expectation value, the gauge group is broken as $so(7) \rightarrow so(5) \times u(1)$. Along this branch, the spinor matters are all massive and integrated out from the low-energy dynamics while the vector matters reduce to the massless $SO(5)$ vectors which can make the vacuum of the low-energy $SO(5)$ theory stable. As a result, the flat direction labeled by $Y$ can become a quantum flat direction. 
Along the $Z$ branch, on the other hand, the gauge group is  broken as $so(7) \rightarrow so(3) \times su(2) \times u(1)$. The massless components of the vector matter make only the $SO(3)$ dynamics stable while the massless components of the spinor matters can make both the $SO(3) \times SU(2)$ dynamics stable. Therefore, the $Spin(7)$ theory with spinor matters can have this flat direction.

The Higgs branch is the same as the 4d case. There are five composite operators
\begin{gather*}
M_{QQ}:=QQ,  \qquad M_{SS}:=SS \\
P_{A1}:=SAS,  \qquad P_{A2}:= SQ^2S,  \qquad P_{S3}:=SQ^3S,
\end{gather*}
where the spinor flavor indices are anti-symmetrized in $P_{A1}, P_{A2}$ and symmetrized in $M_{SS}, P_{S3}$. The quantum numbers of the matter contents and the moduli coordinates are summarized in Table \ref{Spin(7)electric}.  

\begin{table}[H]\caption{3d $\mathcal{N}=2$ $Spin(7)$ with $(N_v,N_s)=(3,3)$} 
\begin{center}
\scalebox{1}{
  \begin{tabular}{|c||c||c|c|c|c|c| } \hline
  &$Spin(7)$&$SU(3)$&$SU(3)$&$U(1)$&$U(1)$&$U(1)_R$  \\ \hline
$Q$&$ \mathbf{7}$ &${\tiny \yng(1)}$&1&$1$&0&$r_v$  \\ 
$S$&$\mathbf{8}$&1&${\tiny \yng(1)}$&$0$&$1$&$r_s$ \\ \hline
$M_{QQ}:=QQ$&1&${\tiny \yng(2)}$&1&2&0&$2r_v$ \\
$M_{SS}:=SS$&1&1&${\tiny \yng(2)}$&0&2&$2r_s$ \\
$P_{A1}:=SQS$&1&${\tiny \yng(1)}$&${\tiny \overline{\yng(1)}}$&1&2&$r_v+2r_s$ \\
$P_{A2}:=SQ^2S$&1&${\tiny  \overline{\yng(1)}}$&${\tiny \overline{\yng(1)}}$&2&2&$2r_v+2r_s$ \\
$P_{S3}:=SQ^3S$&1&1&${\tiny  \yng(2)}$&3&2&$3r_v+2r_s$ \\  \hline
$Y :=\sqrt{Y_1^2 Y_2^2 Y_3}$&1&1&1&$-3$&$-6$&$4-3r_v-6r_s$ \\
$Z:=Y_1 Y_2^2 Y_3$&1&1&1&$-6$&$-6$&$4-6r_v-6r_s$ \\ \hline
  \end{tabular}}
  \end{center}\label{Spin(7)electric}
\end{table}

We consider the dual description of the $Spin(7)$ with $(N_v,N_s)=(3,3)$. The magnetic theory is again the 3d $\mathcal{N}=2$ $Spin(7)$ theory with three vectors and three spinors. In addition, the dual theory includes all the moduli coordinates of the electric theory as elementary fields except for $M_{QQ}$. The tree-level superpotential takes
\begin{align}
W= M_{SS} sq^3s + P_{A1} sq^2s +P_{A2} sqs +P_{S3} ss + Y \tilde{Z} +Z \tilde{Y},
\end{align}
which is consistent with all the symmetries in Table \ref{SO(7)dual}. The above superpotential lifts almost all the flat directions of the magnetic theory and the gauge invariant operator coincide with the electric side. Only the non-trivial identification is the vector meson $M_{QQ}:=QQ \sim qq$. The magnetic Coulomb branch is also two-dimensional and parametrized by $\tilde{Y}, \tilde{Z}$. These are excluded from the chiral ring by the F-flatness conditions of $Z$ and $Y$ respectively.

\begin{table}[H]\caption{Self-dual of $Spin(7)$ with $(N_v,N_s)=(3,3)$} 
\begin{center}
\scalebox{1}{
  \begin{tabular}{|c||c||c|c|c|c|c| } \hline
  &$Spin(7)$&$SU(3)$&$SU(3)$&$U(1)$&$U(1)$&$U(1)_R$  \\ \hline
$q$&$ \mathbf{7}$ &${\tiny \yng(1)}$&1&$1$&0&$r_v$  \\ 
$s$&$\mathbf{8}$&1&${\tiny \overline{\yng(1)}}$&$-\frac{3}{2}$&$-1$&$1-\frac{3}{2}r_v -r_s$ \\
$M_{SS}$&1&1&${\tiny \yng(2)}$&0&2&$2r_s$ \\
$P_{A1}$&1&${\tiny \yng(1)}$&${\tiny \overline{ \yng(1)}}$&1&2&$r_v+2r_s$ \\
$P_{A2}$&1&${\tiny  \overline{\yng(1)}}$&${\tiny \overline{\yng(1)}}$&2&2&$2r_v+2r_s$ \\
$P_{S3}$&1&1&${\tiny  \yng(2)}$&3&2&$3r_v+2r_s$ \\   
$Y $&1&1&1&$-3$&$-6$&$4-3r_v-6r_s$ \\
$Z$&1&1&1&$-6$&$-6$&$4-6r_v-6r_s$ \\ \hline
$M_{QQ} \sim qq$&1&${\tiny \yng(2)}$&1&2&0&$2r_v$ \\  
$ss$&1&1&${\tiny \overline{\yng(2)}}$&$-3$&$-2$&$2-3r_v -2r_s$ \\
$sqs$&1&${\tiny \yng(1)}$&${\tiny \yng(1)}$&$-2$&$-2$&$2-2r_v-r_s$ \\
$sq^2s$&1&${\tiny \overline{\yng(1)}}$&${\tiny \yng(1)}$&$-1$&$-2$&$2 -r_v-2r_s$ \\
$sq^3s$&1&1&${\tiny \overline{\yng(2)}}$&0&$-2$&$2-2r_s$ \\ \hline
$\tilde{Y}$&1&1&1&$6$&$6$&$-2 +6r_v +6r_s$ \\
$\tilde{Z}$&1&1&1&3&6&$-2 +3r_v+r_s$ \\ \hline
  \end{tabular}}
  \end{center}\label{SO(7)dual}
\end{table}

Several consistency checks of the $Spin(7)$ self-duality can be performed. First, the parity anomaly matching is satisfied. By giving the vev to one vector, the electric theory flows to the 3d $\mathcal{N}=2$ $SU(4)$ gauge theory with $2 \, {\tiny \yng(1,1)}+ 3({\tiny \yng(1)}+{\tiny \overline{\yng(1)}})$. The dual gauge group is also broken to $SU(4)$. The magnetic superpotential is decomposed into \eqref{WSU4} and corresponds to the self-dual description studied in Section \ref{SU(4)}. We can also check that these two descriptions exhibit the same superconformal indices \cite{Bhattacharya:2008bja, Kim:2009wb, Imamura:2011su, Kapustin:2011jm}. The index is computed as

\scriptsize
\begin{align}
I_{Spin(7)} &=1+6 \sqrt{x} \left(t^2+u^2\right)+9 t u^2 x^{3/4}+x \left(\frac{1}{t^6 u^6}+21 t^4+45 t^2 u^2+21 u^4\right) \nonumber \\
&\quad +x^{5/4} \left(60 t^3 u^2+54 t u^4\right)+x^{3/2} \left(\frac{6}{t^6 u^4}+56 t^6+\frac{6}{t^4 u^6}+180 t^4 u^2+225 t^2 u^4+56 u^6\right) \nonumber \\
&\quad +x^{7/4} \left(\frac{9}{t^5 u^4}+225 t^5 u^2+\frac{1}{t^3 u^6}+435 t^3 u^4+189 t u^6\right) \nonumber \\
&\qquad+x^2 \left(\frac{1}{t^{12} u^{12}}+126 t^8+525 t^6 u^2+\frac{21}{t^6 u^2}+1125 t^4 u^4+\frac{45}{t^4 u^4}+795 t^2 u^6+\frac{21}{t^2 u^6}+126 u^8-18\right)+\cdots,
\end{align}
\normalsize

\noindent where $t$ is a fugacity for the vector $U(1)$ symmetry and $u$ is a fugacity for the spinor $U(1)$ symmetry. The r-charges are set to $r_v=r_s =\frac{1}{4}$ for simplicity. The low-lying operators corresponds to the moduli coordinates listed in Table \ref{Spin(7)electric}. The Coulomb branch $Z$ corresponds to $\frac{x}{t^6 u^6}$ while $Y$ appears as $\frac{x^{7/4}}{t^3 u^6}$. The higher order terms are fermion contributions and the symmetric products of the moduli operators.

\subsection{$Spin(8)$ self-duality}

\subsubsection{$(N_v, N_s, N_{s'}) =(4,3,0)$}

Let us move on to the self-duality in the 3d $\mathcal{N}=2$ $Spin(8)$ gauge theory with four vectors and three spinors without a tree-level superpotential. The Coulomb branch of the $Spin(8)$ theory was studied in \cite{Nii:2018wwj}. The Higgs branch of the $Spin(8)$ theory is identical to the 4d theory analysis \cite{Csaki:1997cu} with reduction of $N_s$ from four to three. We need to introduce the four Higgs branch operators
\begin{gather*}
M_{QQ} :=QQ, \qquad M_{SS}:=SS  \\
P_{A2} :=SQ^2S,   \qquad P_{S4} :=SQ^4S,
\end{gather*}
where the spinor flavor indices of $M_{SS}, P_{S4}$ are symmetrized and the indices of $P_{A2}$ are anti-symmetric. Table \ref{Spin(8)electric} summarizes the quantum numbers of these operators.

The Coulomb branch is now two-dimensional and parametrized by $Y$ and $Z$ \cite{Nii:2018wwj}. The first Coulomb branch $Y$ corresponds to the breaking $so(8)  \rightarrow so(6) \times u(1)$ and the second Coulomb branch corresponds to $so(8) \rightarrow so(4) \times su(2) \times u(1)$. Along the $Y$ branch, all the components of the spinor matters are massive while the vector matters reduce to the massless $SO(6)$ vectors which make the vacuum of the low-energy $SO(6)$ theory stable. Along the $Z$ branch, the massless components of the vector and spinor matters can make the vacuum of the low-energy $SO(4) \times SU(2)$ theory stable.
The quantum numbers of the Coulomb branch coordinates are shown in Table \ref{Spin(8)electric}.

\begin{table}[H]\caption{3d $\mathcal{N}=2$ $Spin(8)$ with $(N_v,N_s,N_{c})=(4,3,0)$} 
\begin{center}
\scalebox{1}{
  \begin{tabular}{|c||c||c|c|c|c|c| } \hline
  &$Spin(8)$&$SU(4)$&$SU(3)$&$U(1)$&$U(1)$&$U(1)_R$  \\ \hline
$Q$&$ \mathbf{8_v}$ &${\tiny \yng(1)}$&1&$1$&0&$r_v$  \\ 
$S$&$\mathbf{8_s}$&1&${\tiny \yng(1)}$&$0$&$1$&$r_s$ \\ \hline
$M_{QQ}:=QQ$&1&${\tiny \yng(2)}$&1&2&0&$2r_v$ \\
$M_{SS}:=SS$&1&1&${\tiny \yng(2)}$&0&2&$2r_s$ \\
$P_{A2}:=SQ^2S$&1&${\tiny \yng(1,1)}$&${\tiny \overline{\yng(1)}}$&2&2&$2r_v+2r_s$ \\
$P_{S4}:=SQ^4S$&1&1&${\tiny  \yng(2)}$&4&2&$4r_v+2r_s$ \\  \hline
$Y :=\sqrt{Y_1^2 Y_2^2 Y_3Y_4}$&1&1&1&$-4$&$-6$&$4-4r_v-6r_s$ \\
$Z:=Y_1 Y_2^2 Y_3Y_4$&1&1&1&$-8$&$-6$&$4-8r_v-6r_s$ \\ \hline
  \end{tabular}}
  \end{center}\label{Spin(8)electric}
\end{table}

The magnetic theory is again the 3d $\mathcal{N}=2$ $Spin(8)$ gauge theory with four vectors and three spinors. The difference is that the magnetic description includes the five gauge singlet chiral superfields, $M_{SS}, P_{A2}, P_{S4}, Y$ and $Z$ with the tree-level superpotential
\begin{align}
W=M_{SS}sq^4s +P_{A2} sq^2s +P_{S4} ss +Z \tilde{Y} +Y \tilde{Z},
\end{align}
which lifts up various flat directions of the magnetic theory.
The quantum numbers of the magnetic fields and the moduli coordinates are summarized in Table \ref{Spin(8)magnetic}. The charge assignment is determined by the above superpotential.

The operator matching is clear from Table \ref{Spin(8)electric} and Table \ref{Spin(8)magnetic}. Almost all the moduli fields are introduced as elementary fields on the magnetic side. The meson constructed from the vector matters is identified with
\begin{align}
M_{QQ}:=QQ  \sim qq.
\end{align}
The Coulomb branch operators $\tilde{Y}, \tilde{Z}$ of the magnetic theory is all lifted by the above superpotential. 

Let us check the validity of this self-duality. First, the parity anomaly matching is satisfied. By applying the self-duality twice, we recover the electric theory. By giving the vev to a single vector field, we can flow to the $Spin(7)$ self-duality which was discussed in the previous subsection.  We can also test the superconformal indices of these two $Spin(8)$ theories. The two theories give us the same index as

\scriptsize
\begin{align}
I_{Spin(8)}^{(4,3,0)} &=1+\sqrt{x} \left(\frac{1}{t^8 u^6}+10 t^2+6 u^2\right)+x \left(\frac{1}{t^{16} u^{12}}+\frac{6}{t^8 u^4}+\frac{10}{t^6 u^6}+55 t^4+78 t^2 u^2+21 u^4\right)  \nonumber \\
&\qquad +x^{3/2} \left(\frac{1}{t^{24} u^{18}}+\frac{6}{t^{16} u^{10}}+\frac{10}{t^{14} u^{12}}+\frac{21}{t^8 u^2}+\frac{78}{t^6 u^4}+220 t^6+\frac{56}{t^4 u^6}+516 t^4 u^2+318 t^2 u^4+56 u^6\right)  \nonumber \\
&\qquad +x^2 \left(\frac{1}{t^{32} u^{24}}+\frac{6}{t^{24} u^{16}}+\frac{10}{t^{22} u^{18}}+\frac{21}{t^{16} u^8}+\frac{78}{t^{14} u^{10}}+\frac{56}{t^{12} u^{12}}+715 t^8+\frac{56}{t^8}+2370 t^6 u^2 \right. \nonumber \\
& \qquad \qquad   \left. +\frac{318}{t^6 u^2}  +2436 t^4 u^4+\frac{516}{t^4 u^4}+938 t^2 u^6+\frac{230}{t^2 u^6}+126 u^8-25\right)+\cdots,
\end{align}
\normalsize 

\noindent where $t$ and $u$ are the fugacities for the vector and spinor $U(1)$ symmetries, respectively. We set the r-charges to $r_v=r_s =\frac{1}{4}$ for simplicity. The second term corresponds to $Z$, $M_{QQ}$ and $M_{SS}$. The third term consists of the symmetric products of these three fields and $P_{A2}$. $P_{S4}$ appears as $t^4 u^2 x^{3/2}$ while $Y$ corresponds to $\frac{x^{3/2}}{t^4 u^6}$.

\begin{table}[H]\caption{Self-dual of $Spin(8)$ with $(N_v,N_s,N_{c})=(4,3,0)$} 
\begin{center}
\scalebox{1}{
  \begin{tabular}{|c||c||c|c|c|c|c| } \hline
  &$Spin(8)$&$SU(4)$&$SU(3)$&$U(1)$&$U(1)$&$U(1)_R$  \\ \hline
$q$&$ \mathbf{8_v}$ &${\tiny \yng(1)}$&1&$1$&0&$r_v$  \\ 
$s$&$\mathbf{8_s}$&1&${\tiny \overline{\yng(1)}}$&$-2$&$-1$&$1-2r_v-r_s$ \\ 
$M_{SS}$&1&1&${\tiny \yng(2)}$&0&2&$2r_s$ \\
$P_{A2}$&1&${\tiny \yng(1,1)}$&${\tiny \overline{\yng(1)}}$&2&2&$2r_v+2r_s$ \\
$P_{S4}$&1&1&${\tiny  \yng(2)}$&4&2&$4r_v+2r_s$ \\  
$Y $&1&1&1&$-4$&$-6$&$4-4r_v-6r_s$ \\
$Z$&1&1&1&$-8$&$-6$&$4-8r_v-6r_s$ \\ \hline
$M_{QQ} \sim qq$&1&${\tiny \yng(2)}$&1&2&0&$2r_v$ \\
$ss$&1&1&${\tiny \overline{\yng(2)}}$&$-4$&$-2$&$2 -4r_v -2r_s$ \\
$sq^2s$&1&${\tiny \yng(1,1)}$&${\tiny \yng(1)}$&$-2$&$-2$&$2-2r_v-2r_s$\\
$sq^4s$&1&1&${\tiny \overline{\yng(2)}}$&0&$-2$&$2 -2r_s$ \\ \hline
$\tilde{Y}$&1&1&1&$8$&6&$-2 +8r_v +6r_s$ \\
$\tilde{Z}$&1&1&1&4&6&$-2 +4r_v +6r_s$ \\ \hline
  \end{tabular}}
  \end{center}\label{Spin(8)magnetic}
\end{table}

\subsubsection{$(N_v, N_s, N_{s'}) =(4,2,1)$}
Finally, we study the self-duality of the 3d $\mathcal{N}=2$ $Spin(8)$ gauge theory with four vectors, two spinors and a single conjugate spinor. The Higgs branch is described by the following composite operators.
\begin{gather*}
M_{QQ}:=QQ,  \quad  M_{SS}:=SS, \quad M_{S'S'}:=S'S' \\
P_1:=SQS',  \quad   P_{2A}:=SQ^2S,  \quad    P_3:=SQ^3S'  \\
P_{4S}:=SQ^4S,  \qquad P_{4C}:=S'Q^4S'
\end{gather*}
The quantum numbers of these operators are summarized in Table \ref{Spin(8)421electric}. The Coulomb branch is the same as the previous case and two-dimensional. The $Y$ coordinate corresponds to the breaking $so(8)  \rightarrow so(6) \times u(1)$ while the $Z$ coordinate is related to the breaking $so(8) \rightarrow so(4) \times su(2) \times u(1)$. These flat directions are made stable and supersymmetric by massless components of the vector and spinor matters.
The quantum numbers of the corresponding operators are also summarized in Table \ref{Spin(8)421electric}.

\begin{table}[H]\caption{3d $\mathcal{N}=2$ $Spin(8)$ with $(N_v,N_s,N_{c})=(4,2,1)$} 
\begin{center}
\scalebox{1}{
  \begin{tabular}{|c||c||c|c|c|c|c|c| } \hline
  &$Spin(8)$&$SU(4)$&$SU(2)$&$U(1)$&$U(1)$&$U(1)$&$U(1)_R$  \\ \hline
$Q$&$ \mathbf{8_v}$ &${\tiny \yng(1)}$&1&1&0&0&$r_v$  \\
$S$&$ \mathbf{8_s}$ &1&${\tiny \yng(1)}$&0&1&0&$r_s$ \\
$S'$&$ \mathbf{8_c}$ &1&1&0&0&1&$r_c$ \\  \hline
$M_{QQ}:=QQ$&1&${\tiny \yng(2)}$&1&2&0&0&$2r_v$ \\
$M_{SS}:=SS$&1&1&${\tiny \yng(2)}$&0&$2$&0&$2r_s$ \\
$M_{S'S'}:=S'S'$&1&1&1&0&0&2&$2r_c$  \\
$P_1:=SQS'$&1&${\tiny \yng(1)}$&${\tiny \yng(1)}$&1&1&1&$r_v+r_s+r_c$ \\
$P_{2A}:=SQ^2S$&1&${\tiny \yng(1,1)}$&1&2&2&0&$2r_v+2r_s$  \\
$P_3:=SQ^3S'$&1&${\tiny \overline{\yng(1)}}$&${\tiny \yng(1)}$&3&1&1&$3r_v+r_s+r_c$  \\
$P_{4S}:=SQ^4S$&1&1&${\tiny \yng(2)}$&4&2&0&$4r_v+2r_s$  \\
$P_{4C}:=S'Q^4S'$&1&1&1&4&0&2&$4r_v+2r_c$  \\ \hline
$Y$&1&1&1&$-4$&$-4$&$-2$&$4-4r_v-4r_s-2r_c$  \\
$Z$&1&1&1&$-8$&$-4$&$-2$&$4-8r_v-4r_s-2r_c$  \\  \hline
  \end{tabular}}
  \end{center}\label{Spin(8)421electric}
\end{table}

Let us consider the self-dual description. The dual theory is again the 3d $\mathcal{N}=2$ $Spin(8)$ gauge theory with four vectors, two spinors and a single conjugate spinor.  In addition, the dual theory includes the elementary gauge singlet fields, $M_{SS}, M_{S'S'}, P_1, P_{2A}, P_3, P_{4S}, P_{4C}, Y$ and $Z$ with the tree-level superpotential
\begin{align}
W= M_{SS} sq^4s + M_{S'S'} s'q^4s' +P_1 sq^3s' + P_{2A} sq^2s +P_3 sqs'+ P_{4S} ss +P_{4C}s's'+Y \tilde{Z} +Z \tilde{Y}.
\end{align}
Therefore, on the dual side, all the moduli coordinates are introduced as elementary fields except for the meson constructed from the vector matters. The non-trivial matching of the chiral ring generators is $M_{QQ}:=QQ \sim qq $. The charge assignment of the dual fields is determined by the superpotential and the meson matching. The quantum numbers of the dual moduli fields are summarized in Table \ref{Spin(8)421magnetic}. 

We can check several consistencies of this self-duality. First, the parity anomaly matching is satisfied. By applying the self-duality twice, we can go back to the electric description. By giving a vacuum expectation value to a single vector, the electric and magnetic theories flow to the $Spin(7)$ duality which was studied in the previous subsections. Finally, we can test the superconformal indices of this self-duality. The indices of the electric and magnetic theories become  

\scriptsize
\begin{align}
I_{Spin(8)}^{(4,2,1)} &=1+\sqrt{x} \left(\frac{1}{t^8 u^4 v^2}+10 t^2+3 u^2+v^2\right)+8 t u v x^{3/4} \nonumber \\
&+x \left(\frac{1}{t^{16} u^8 v^4}+55 t^4+36 t^2 u^2+10 t^2 v^2+\frac{10 t^2+3 u^2+v^2}{t^8 u^4 v^2}+6 u^4+3 u^2 v^2+v^4\right) \nonumber \\
&+x^{5/4} \left(\frac{8}{t^7 u^3 v}+88 t^3 u v+24 t u^3 v+8 t u v^3\right) +x^{3/2} \left(\frac{1}{t^{24} u^{12} v^6}+220 t^6+228 t^4 u^2+56 t^4 v^2+78 t^2 u^4+72 t^2 u^2 v^2  \right.  \nonumber \\
&\left.+10 t^2 v^4+\frac{10 t^2+3 u^2+v^2}{t^{16} u^8 v^4}+\frac{56 t^4+36 t^2 u^2+10 t^2 v^2+6 u^4+3 u^2 v^2+v^4}{t^8 u^4 v^2}+10 u^6+6 u^4 v^2+3 u^2 v^4+v^6\right) \nonumber \\
&+x^{7/4} \left(\frac{8}{t^{15} u^7 v^3}+520 t^5 u v+312 t^3 u^3 v+88 t^3 u v^3+\frac{8 \left(11 t^2+3 u^2+v^2\right)}{t^7 u^3 v}+48 t u^5 v+24 t u^3 v^3+8 t u v^5\right)+\cdots,
\end{align}

\normalsize
\noindent where $t,u$ and $v$ are the fugacities corresponding to the three $U(1)$ global symmetries. We set the r-charges to $r_v=r_s= r_c=\frac{1}{4}$ for simplicity. The Coulomb branch $Z$ is represented as $\frac{x^{\frac{1}{2}}}{t^8 u^4 v^2}$ while $Y$ appears as $\frac{x^{\frac{3}{2}}}{t^4 u^4 v^2}$. The other terms are identified with the Higgs branch operators and the symmetric products of the gauge invariant operators in Table \ref{Spin(8)421electric}.

\begin{table}[H]\caption{Self-dual of $Spin(8)$ with $(N_v,N_s,N_{c})=(4,2,1)$} 
\begin{center}
\scalebox{1}{
  \begin{tabular}{|c||c||c|c|c|c|c|c| } \hline
  &$Spin(8)$&$SU(4)$&$SU(2)$&$U(1)$&$U(1)$&$U(1)$&$U(1)_R$  \\ \hline
$q$&$ \mathbf{8_v}$ &${\tiny \yng(1)}$&1&1&0&0&$r_v$  \\
$s$&$ \mathbf{8_s}$ &1&${\tiny \yng(1)}$&$-2$&$-1$&0&$1-2r_v-r_s$ \\
$s'$&$ \mathbf{8_c}$ &1&1&$-2$&0&$-1$&$1-2r_v -r_c$ \\  
$M_{SS}$&1&1&${\tiny \yng(2)}$&0&$2$&0&$2r_s$ \\
$M_{S'S'}$&1&1&1&0&0&2&$2r_c$  \\
$P_1$&1&${\tiny \yng(1)}$&${\tiny \yng(1)}$&1&1&1&$r_v+r_s+r_c$ \\
$P_{2A}$&1&${\tiny \yng(1,1)}$&1&2&2&0&$2r_v+2r_s$  \\
$P_3$&1&${\tiny \overline{\yng(1)}}$&${\tiny \yng(1)}$&3&1&1&$3r_v+r_s+r_c$  \\
$P_{4S}$&1&1&${\tiny \yng(2)}$&4&2&0&$4r_v+2r_s$  \\
$P_{4C}$&1&1&1&4&0&2&$4r_v+2r_c$  \\
$Y$&1&1&1&$-4$&$-4$&$-2$&$4-4r_v-4r_s-2r_c$  \\
$Z$&1&1&1&$-8$&$-4$&$-2$&$4-8r_v-4r_s-2r_c$  \\  \hline
$M_{QQ} \sim qq$&1&${\tiny \yng(2)}$&1&2&0&0&$2r_v$ \\  \hline 
$\tilde{Y}$&1&1&1&$8$&4&2&$-2 +8r_v +4r_s+2r_c$  \\
$\tilde{Z}$&1&1&1&$4$&4&2&$-2 +4r_v +4r_s+2r_c$  \\  \hline
  \end{tabular}}
  \end{center}\label{Spin(8)421magnetic}
\end{table}

\section{Summary and Discussion}
In this paper, we constructed the self-dualities in 3d $\mathcal{N}=2$ supersymmetric gauge theories with various gauge groups and various matter contents. The examples include the $SU(2)$, $SU(4)$, $SU(6)$, $SU(8)$, $SU(2N)$, $Spin(7)$ and $Spin(8)$ gauge theories. The 3d self-dualities are related to the 4d self-dualities via dimensional reduction and a real mass deformation. In 3d, there is an additional moduli space called the Coulomb branch and the matching of the moduli coordinates becomes complicated, compared to the 4d case. The (dressed) Coulomb branch is mapped to the (anti-)baryon operator, to the gauge singlet chiral superfields or to the magnetic Coulomb branch. For some self-dual examples, we computed the superconformal indices of the self-dual pair and observed the agreement of the indices. For the $SU(6)$ case, we also derived the duality between the $USp(4)$ and $SU(6)$ gauge theories, which exhibited the symmetry enhancement.

It is worth further checking the validity of our self-dualities. For example, one can compute the partition function on both sides of the self-dualities. This would be an independent check of the self-dualities. 
It would be important to study the connection between the 3d and 4d self-dualities by using the 4d superconformal indices and the 3d partition function, which are related to each other via dimensional reduction. Although we found the relation between the 3d and 4d self-dualities, the mechanism of the dynamical generation of the monopole potential like \eqref{WSU4} was unclear. We will leave this problem to a future direction. For the $Spin(N)$ self-dualities, we could not find the relation between the 3d and 4d $Spin(N)$ self-dualities. This would be also left as a future problem.
Although we constructed the various self-dualities in this paper, we did not exhaust all the possibilities of the 3d self-dualities. It is quite important to search for other 3d self-dualities, especially for the $Spin(N)$ cases with vector and spinor matters.

\section*{Acknowledgments}
I would like to thank Antonio Amariti for valuable discussions.
This work is supported by the Swiss National Science Foundation (SNF) under grant number PP00P2\_157571/1 and by ``The Mathematics of Physics'' (SwissMAP) under grant number NCCR 51NF40-141869.


\bibliographystyle{ieeetr}
\bibliography{3dSelfDual}

\end{document}